\documentclass[aps,prd,reprint,twocolumn,preprintnumbers,floatfix,nofootinbib]{revtex4-1}
\usepackage[utf8]{inputenc}
\pdfoutput=1
\usepackage{graphicx}
\usepackage{amsmath}
\usepackage{amsfonts}
\usepackage{amssymb}
\usepackage{xcolor,soul}
\usepackage{epstopdf}
\usepackage{xcolor}
\usepackage{float}
\usepackage{subfigure}
\usepackage{hyperref}
\usepackage[hyphenbreaks]{breakurl}
\usepackage[left=2cm,right=2cm,top=2cm,bottom=2cm]{geometry}
\usepackage{relsize}
\usepackage{graphics}
\usepackage{hyperref}
\usepackage{mathrsfs}
\usepackage{amssymb}
\usepackage{booktabs}
\usepackage[normalem]{ulem}
\usepackage{amsthm}
\usepackage{cancel}
\usepackage{fontawesome}   
\usepackage{tcolorbox}
\usepackage{tabularx,ragged2e} 
\usepackage[nodisplayskipstretch]{setspace}

\hypersetup{
     colorlinks   = true,
     citecolor    = red,
     linkcolor    = blue,
     urlcolor     = blue,
}
\usepackage{physics}  
\newcommand{\bea}{\begin{aligned}}
\newcommand{\eea}{\end{aligned}}
\def\beq{\begin{equation}}
\def\eeq{\end{equation}}
\def\beqa{\begin{eqnarray}}
\def\eeqa{\end{eqnarray}}
\def\be{\begin{equation}}
\def\ee{\end{equation}}

\def\bse{\begin{subequations}}
\def\ese{\end{subequations}}

\def\trh{$T_{\mathrm{RH}}~$}

\def\tin{t_{\rm in}}

\def\ain{a_{\rm in}}
\def\ae{a_{\mathrm{end}}}
\def\arh{a_{\mathrm{RH}}}
\def\abh{a_{\mathrm{BH}}}

\def\aev{a_{\rm ev}}
\def\rhorh{\rho_{\mathrm{RH}}}
\def\trh{T_{\mathrm{RH}}}
\def\bea{\begin{eqnarray}}
\def\eea{\end{eqnarray}}
\def\rhoe{\rho_{\mathrm{end}}}
\def\ai{a_{\mathrm{in}}}
\def\aev{a_{\mathrm{ev}}}
\def\Min{M_{\mathrm{in}}}
\def\BH{\mathrm{BH}}

\def\amax{a_{\rm max}}
\def\rhobh{\rho_{\rm BH}}

\newcommand{\rbh}{\rho_{\rm BH}}
\newcommand{\rphi}{\rho_{\phi}}
\newcommand{\Tev}{T_\text{ev}}

\newcommand{\Tbh}{T_\text{BH}}
\newcommand{\nbh}{n_\text{BH}}
\newcommand{\krh}{k_\text{RH}}
\newcommand{\Hrh}{H_\text{RH}}

\newcommand{\kbh}{k_\text{BH}}
\newcommand{\kmax}{k_\text{max}}
\newcommand{\DNeff}{\Delta N_{\rm eff}}

\newcommand{\aend}{a_{\rm end}}

\graphicspath{{./figs/}}
\usepackage{pgfplots}
\pgfplotsset{compat=1.17}
\begin{document}
\preprint{CTPU-PTC-24-07}
\vspace*{1mm}
\title{Leptogenesis, primordial gravitational waves, and PBH-induced
    reheating}
\author{Basabendu Barman$^{a}$}
\email{basabendu.b@srmap.edu.in}
\author{Suruj Jyoti Das$^{b}$}
\email{surujjd@gmail.com}
\author{Md Riajul Haque$^{c}$}
\email{riaj.0009@gmail.com}
\author{Yann Mambrini$^{d}$}
\email{yann.mambrini@ijclab.in2p3.fr}
\vspace{0.1cm}
\affiliation{
${}^a$
Department of Physics, School of Engineering and Sciences,
SRM University-AP, Amaravati 522240, India}
\affiliation{
${}^b$
Particle Theory  and Cosmology Group, Center for Theoretical Physics of the Universe,
Institute for Basic Science (IBS),
 Daejeon, 34126, Korea
}
\affiliation{
${}^c$
Centre for Strings, Gravitation, and Cosmology,
Department of Physics, Indian Institute of Technology Madras, 
Chennai~600036, India
}
\affiliation{
${}^d$ Universit\'e Paris-Saclay, CNRS/IN2P3, IJCLab, 91405 Orsay, France
}
\begin{abstract} 
We explore the possibility of 
producing the observed matter-antimatter asymmetry of the Universe uniquely from the evaporation of primordial black holes (PBH) that are formed in an inflaton-dominated 
background. Considering the inflaton $(\phi)$ to oscillate in a monomial 
potential $V(\phi)\propto\phi^n$, we show, it is possible to obtain 
the desired baryon asymmetry via vanilla leptogenesis from evaporating 
PBHs of initial mass $\lesssim 10$ g. 
We find that the allowed parameter space is 
heavily dependent on the shape of the inflaton potential during reheating 
(determined by the exponent of the potential $n$), the energy density of PBHs (determined by 
$\beta$), and the nature of the coupling between the inflaton and the 
Standard Model (SM). To complete the minimal gravitational framework, we also include in our analysis the
gravitational leptogenesis set-up through inflaton scattering via exchange of graviton, which opens up an even larger window for PBH mass, depending on the background equation of state.  We finally illustrate that such gravitational 
leptogenesis scenarios can be tested with upcoming gravitational wave (GW) 
detectors, courtesy of the blue-tilted primordial GW with 
inflationary origin, thus paving a way to probe a PBH-induced reheating together with leptogenesis. 
\end{abstract}
\maketitle
{
  \hypersetup{linkcolor=black}
  \tableofcontents
}
\section{Introduction}
\label{sec:intro}
Initially proposed by Stephen Hawking and Bernard Carr, primordial black holes (PBHs) exhibit captivating cosmic signatures~\cite{Hawking:1974rv, Hawking:1975vcx}. PBHs with masses  $M_{\rm BH} \gtrsim 10^{15}$ g 
remain stable at the present day and can be suitable for dark matter (DM) candidates (see, for example, Ref.~\cite{Carr:2020xqk} for a review). 
On the other side of the spectrum, 
the black holes must be much lighter to explore particle production from evaporating PBHs. 
Indeed, the formation mass should be within a range allowing for evaporation before Big Bang Nucleosynthesis (BBN),
corresponding to $M_{\rm BH} \lesssim 10^{9}$ g. 
Failure to meet this criterion could introduce additional degrees of freedom, potentially disrupting the successful prediction of BBN from the accurate measurement of $\DNeff$~\cite{Planck:2018vyg}. 
Within this mass range, PBHs can undergo decay and 
play a central role in producing Standard Model (SM) particles, DM, and baryon asymmetry. Various studies that have explored DM production~\cite{Morrison:2018xla, Gondolo:2020uqv, Bernal:2020bjf, Green:1999yh, Khlopov:2004tn, Dai:2009hx, Allahverdi:2017sks, Lennon:2017tqq, Hooper:2019gtx, Chaudhuri:2020wjo, Masina:2020xhk, Baldes:2020nuv, Bernal:2020ili, Bernal:2020kse, Lacki:2010zf, Boucenna:2017ghj, Adamek:2019gns, Carr:2020mqm, Masina:2021zpu, Bernal:2021bbv, Bernal:2021yyb, Samanta:2021mdm, Sandick:2021gew, Cheek:2021cfe, Cheek:2021odj, Barman:2021ost, Borah:2022iym,Chen:2023lnj,Chen:2023tzd,Kim:2023ixo,Gehrman:2023qjn}, 
baryon asymmetry~\cite{Hawking:1974rv, Carr:1976zz,Baumann:2007yr, Hook:2014mla, Fujita:2014hha, Hamada:2016jnq, Morrison:2018xla, Hooper:2020otu, Perez-Gonzalez:2020vnz, Datta:2020bht, JyotiDas:2021shi, Smyth:2021lkn, Barman:2021ost, Bernal:2022pue, Ambrosone:2021lsx,Calabrese:2023key,Calabrese:2023bxz,Gehrman:2022imk,Gehrman:2023esa} or cogenesis~\cite{Fujita:2014hha, Morrison:2018xla, Hooper:2019gtx, Lunardini:2019zob, Masina:2020xhk, Hooper:2020otu, Datta:2020bht, JyotiDas:2021shi, Schiavone:2021imu, Bernal:2021yyb, Bernal:2021bbv, Bernal:2022swt,Barman:2022pdo} from PBH evaporation, have consistently focused on PBH formation during standard radiation domination, overlooking the evolution of PBHs in a cosmological background dominated by the inflaton field \footnote{For the effect of reheating on flavor leptogenesis, see, for instance, Refs. \cite{Datta:2022jic,Datta:2023pav}.}. However, recently, the authors of ~\cite{RiajulHaque:2023cqe,Haque:2024cdh} studied the aftermath of PBH formation and evaporation
{\it during} reheating, in presence 
of the inflaton field. They focused mainly on the 
production of DM relic from Schwarzschild BH and the 
effect of PBH decay on the reheating temperature. 
From these studies it was established that (a) the inflaton decay is more efficient at the beginning of the reheating 
process, whereas the evaporation of PBHs is more efficient at the end of their lifetime, and (b) PBH evaporation in an inflaton-dominated Universe can produce the entire observed DM relic abundance and even dominate the reheating process. 
Combining these two natural sources (PBH and inflaton) of radiation and dark matter reopened a large part of the parameter space, which was forbidden.

If PBHs can greatly influence the production of particles 
in the early Universe still dominated with the inflaton field, 
it is natural to ask about the generation of baryon asymmetry through their evaporation within the epoch of reheating. As we know, an elegant mechanism to produce the baryon asymmetry of the Universe (BAU) is via  leptogenesis~\cite{Fukugita:1986hr,Davidson:2008bu}, where a lepton asymmetry is generated first and subsequently gets converted into baryon asymmetry via non-perturbative sphaleron transitions~\cite{Kuzmin:1985mm}. In standard thermal leptogenesis~\cite{Buchmuller:2002rq,Buchmuller:2003gz,Chankowski:2003rr,Giudice:2003jh}, the decaying particles, typically right-handed neutrinos (RHNs), are produced thermally from the SM bath. However, the lower bound on the RHN mass in such scenarios (known as the Davidson-Ibarra bound), leads to a lower bound on the reheating temperature $\trh\gtrsim 10^{10}$ GeV~\cite{Davidson:2002qv}, leading to the so-called ``gravitino problem"~\cite{PhysRevLett.48.1303,Davidson:2008bu}. A simple alternative to circumvent this, is to consider non-thermal production of RHNs~\cite{Giudice:1999fb,Asaka:1999yd,Lazarides:1990huy,Campbell:1992hd,Hahn-Woernle:2008tsk,Barman:2021tgt}, that can be sourced by the PBH evaporation. 

On the other hand, there also exists an unavoidable 
production of RHNs 
through the gravitational
interaction~\cite{Choi:1994ax,Holstein:2006bh}. 
Indeed, it was shown in~\cite{Co:2022bgh,Barman:2022qgt}
that the transfer of energy from the inflaton background
field can produce RHNs via the exchange of a massless graviton and is
a valid source of BAU\footnote{Such gravitational interaction can also reheat the Universe via gravitational reheating~\cite{Clery:2021bwz,Haque:2022kez,Clery:2022wib}.} and in another possibility, decay of those gravitationally produced RHNs may lead to the radiation dominated universe \cite{Haque:2023zhb}.
Therefore, this coupling being unavoidable makes it impossible to ignore the production of RHNs from the scattering of inflaton condensate, mediated by massless gravitons. 
In the present set-up, we consider both contributions, namely, asymmetry from PBH evaporation and also from 
the graviton-mediated process, trying to combine both sources and find in which part of the parameter space one
of the 
source dominates over the other one.

It is also important to note that for inflaton oscillating in a monomial potential $V(\phi)\propto\phi^n$ where the equation of state (EoS) is given by $w_\phi=(n-2)/(n+2)$, the value $ w_\phi>1/3$ (equivalently, $n>4$) plays a crucial 
role in probing the reheating scenario with primordial GWs (PGW)\footnote{For other relevant sources of PGW see, for example, the recent review~\cite{Roshan:2024qnv}.}, that are originated from the 
tensor fluctuations during inflation. Such a {\it stiff} period in the expansion history significantly enhances the inflationary GW background, 
making the corresponding signal potentially observable at several GW experimental facilities~\cite{Giovannini:1998bp,Giovannini:1999bh,Riazuelo:2000fc,Seto:2003kc,Boyle:2007zx,Stewart:2007fu,Li:2021htg,Artymowski:2017pua,Caprini:2018mtu,Bettoni:2018pbl,Bernal:2019lpc,Figueroa:2019paj,Opferkuch:2019zbd,Bernal:2020ywq,Ghoshal:2022ruy,Caldwell:2022qsj,Gouttenoire:2021jhk,Barman:2023ktz,Chakraborty:2023ocr,Barman:2023icn}. In the present context, the blue-tilted GW spectrum 
turns out to be well within the reach of several future GW detectors. More importantly, the red-
tilted spectrum due to intermediate PBH-domination also turns out to be potentially detectable in 
detectors like BBO~\cite{Crowder:2005nr,Corbin:2005ny} and DECIGO~\cite{Kawamura:2006up,Kawamura:2020pcg}. This paves a way to testability of the 
present scenario, where any future detection can not only validate the inflationary paradigm but 
also hint towards a non-trivial cosmological history of the Universe prior to the onset of BBN. 

The paper is organized as follows. After computing the BAU generated by PBH decay in the presence of 
the inflaton field in section II, we compare it with the asymmetry produced through gravitational 
interaction in section III. We then analyze possible GW signatures of our scenario in section 
IV, before concluding. 

\begin{tcolorbox}[colback=gray!5!white,colframe=gray!50!black,colbacktitle=gray!75!black,title=Guide to scale-factor notations]
\begin{itemize}
\item [] $\aend:$ Scale factor at the end of inflation
\item [] $\ain:$ Scale factor at PBH formation
\item [] $\abh:$ Scale factor at the onset of PBH domination 
\item [] $\aev:$ Scale factor at PBH evaporation 
\item [] $\arh:$ Scale factor at the end of reheating 
\end{itemize}
\end{tcolorbox}

\section{Leptogenesis from PBH} 
\label{sec:lepto}
\subsection{Generalities}
Assuming that PBHs have been formed during the reheating phase, their mass
 is typically related to the energy enclosed in the particle horizon. 
The mass $\Min$ at formation time is given by~\cite{Carr:1974nx}
\beq
\Min=\frac{4}{3}\,\pi\,\gamma\,H_{\rm in}^{-3}\,\rho_\phi(\ain)
=4\,\pi\,\gamma\, M_P^2\,H_{\rm in}^{-1}\,,
\label{Eq:min}
\eeq
where $\gamma=w_\phi^{{3}/{2}}$ parameterizes 
the efficiency of the collapse to form PBHs
and $M_P= 1/\sqrt{8 \pi G_N}\simeq 2.4\times 10^{18}$ GeV is the reduced Planck mass\footnote{We can always choose a formation mechanism other than the horizon collapse, where the formation mass will differ. However, once we fixed the formation mass, the rest of the analysis related to reheating and leptogenesis via PBH-inflaton interplay
remains as it is.}. $\rho_\phi(\ain)$ and $H_{\rm in}$ are 
the inflaton energy density and Hubble parameter, respectively, at the time of 
formation corresponding to the scale factor $a_{\rm in}$. In addition, the PBH mass evolves as (see Eq.~\eqref{eq:beq} in Appendix~\ref{sec:BEQ})
\beq
M_{\rm BH}(t)=\Min\left(1 - \Gamma_{\rm BH} (t-\tin)\right)^{\frac{1}{3}}\,,
\label{Eq:mbht}
\eeq
where 
\beq
\Gamma_{\rm BH}= 3 \,\epsilon \frac{M_P^4}{M_{\rm in}^3}\,,
\label{Eq:gammabh}
\eeq
with $\tin$ being the time of the formation and

\beq
\epsilon=\mathcal{G}\times \frac{\pi\,g_* (T_{\rm BH})}{480}\,,
\label{Eq:epsilon}
\eeq
where $\mathcal{G}=27/4$ is the 
greybody factor~\cite{PhysRevD.41.3052}\footnote{A more comprehensive expression for greybody factor can be found, for example, in Refs.~\cite{Auffinger:2020afu,Masina:2021zpu,Cheek:2021odj}.}. In the case of a Schwarzschild BH, which we will consider in the rest of our analysis, its temperature $\Tbh$ is related to its mass via~\cite{Hawking:1975vcx} 
\begin{align}\label{eq:TBH}
& \Tbh=M_P^2/M_{\rm BH}\,.  
\end{align}
Finally, the PBH mass can not take arbitrary values. 
It is bounded from above and below within the window 
\begin{align}
& 1\,\text{g}\lesssim\Min\lesssim 10^8\,\text{g}\,,    
\end{align}
where the lower bound arises from the size of the horizon at the end of inflation $\Min\gtrsim H_{\rm end}^{-3}\,\rhoe\approx 1\,\text{g}\,\left(10^{64}\,\text{GeV}^4/\rhoe\right)^{1/2}$, while the upper bound emerges by requiring PBH evaporation before the onset of BBN: $t_{\rm ev}\simeq 1\,{\rm sec}\,\left(\Min/10^8\,\text{g}\right)^3$. The time of evaporation $t_{\rm ev}$ can be obtained by solving the PBH mass evolution, Eq.~\eqref{eq:beq}, which corresponds roughly to the condition $M_{\rm BH}=0$ in Eq.~(\ref{Eq:mbht}), or $t_{\rm ev}\simeq \Gamma_{\rm BH}^{-1}$ when $t_{\rm in} \ll t_{\rm ev}$.  

The total number of particles emitted during PBH evaporation depends on its intrinsic properties, such as the particle's spin and mass. The production rate for any species $X$ with internal degrees of freedom $g_X$ can be estimated as~\cite{Masina:2020xhk}
\beq\label{Eq:emitionrate}
\frac{d\mathcal{N}_i}{dt}=\frac{27}{4}\frac{g_X\,\xi\,\zeta(3)}{16 \pi^3}\frac{M_P^2}{M_{\rm BH}(t)}\,,
\eeq
where $\xi=(1,\,3/4)$ for bosonic and fermionic fields, respectively, and $g_X$ is the internal degrees of freedom for the corresponding field. After integration, we obtain

\begin{equation}
    \mathcal{N}_i = \frac{15\,g_X\,\xi\,\zeta(3)}{g_\star(T_\text{BH}^{\rm in})\pi^4}
    \begin{cases}
       \left(\frac{M_{\rm in}}{M_P}\right)^2 \,,~M_X < T_\text{BH}^\text{in}\,,\\[8pt]
        \left(\frac{M_{P}}{M_X}\right)^2 \,,~M_X > T_\text{BH}^\text{in}\,,
    \end{cases}\label{eq:pbh-num}
\end{equation}
where $M_X$ is the mass of the corresponding species and 
\begin{align}
& T_{\rm BH}^{\rm in}=M_P^2/\Min\simeq 10^{13}\left(\frac{1~\rm{g}}{M_{\rm in}}\right)~\rm{GeV}\,,  
\end{align}
is the PBH temperature at the point of formation\footnote{For the mass scale we consider in the work, $g_{\star}=106.75$, 
$T_{\rm BH}$ being much larger than the electroweak scale, all the SM degrees of freedom should be taken into account.}. 

Note that if one considers the production 
of RHNs with mass $M_N<T_\text{BH}^\text{in}$, 
PBH should emit them from the formation time $t_{\rm in}$, whereas for $M_N>T_\text{BH}^\text{in}$, PBH starts to emit RHNs when PBH temperature $T_{\rm BH}\sim M_N$. Out-of-equilibrium production of RHNs 
is a key ingredient in the leptogenesis scenario.
Indeed, SM can be extended by taking three right-handed SM singlet massive neutrino $N_{\rm i}(i=1,\,2\,,3)$ with the interaction lagrangian
\beq
 \mathcal{L} \supset -\frac12\, \sum_i M_{N_i}\, \overline{N_i^c}\, N_i - y_{N}^{ij}\, \overline{N_i}\, \widetilde{H}^\dagger\, L_j + {\rm h.c.}\,,
\eeq
where SM left-handed leptons doublets are identified as $L_i$ and $\widetilde{H}=i\,\sigma_{\rm 2}\,H^{\rm *}$ where $H$ represents the SM Higgs doublet. $\sigma_{\rm i}$ are the Pauli spin matrices. 
We detail the Yukawa coupling parametrization in the Appendix~\ref{sec:CI}. We assume the Majorana masses $M_{N_i}$ to be hierarchical $M_{N_1} \ll M_{N_{2,3}}$. 
Moreover, for the decay of heavier RHNs $N_{2,3}$, we consider lepton-number-violating interactions of $N_1$ rapid enough to wash out the lepton-number 
asymmetry originated by the other two. Therefore, only the CP-violating asymmetry from the decay of $N_1$ survives and is relevant for leptogenesis\footnote{The 
effects due to $N_{2,3}$ can be neglected as long as $\text{max}\,[\trh,\,\Tev]<M_{N_{2,3}}$, which we consider in the present analysis. In the opposite 
case,  $L$-violating interactions of $N_1$ does not wash out any lepton asymmetry generated at temperatures $T\gg M_{N_1}$ via decays of $N_{2,3}$. In such scenarios, 
the lepton asymmetry generated in $N_{2,3}$ decays survives the $N_1$ leptogenesis phase.}.

Once right-handed neutrinos are produced from the evaporating PBHs, they can decay later and produce lepton asymmetry, which can be converted into the baryon asymmetry through the Standard Model electroweak sphaleron process. At the origin
of the asymmetry generation, right-handed neutrino decay rapidly into left-handed lepton $L$ and Higgs doublets $H$, $N\to L+H$, and $N\to \bar{L}+\tilde{H}$ and if CP is violated, the lepton asymmetry is then given by
\beq
Y_L=\frac{n_L}{s}=\kappa_{\Delta L}\,\frac{n_{N_1}}{s}\,,
\eeq
where $s=\frac{2\,\pi^2}{45}\,g_{*}(T)\,T^3$ represents entropy energy density with radiation temperature $T$,  and the CP asymmetry generated from $N_1$ decay is given by~\cite{Davidson:2008bu}
\begin{align}\label{eq:cp-asym}
&\kappa_{\Delta L} \equiv \frac{\Gamma_{N_1 \to \ell_i\, H } -\Gamma_{N_1 \to \bar\ell_i\, \bar H}}{\Gamma_{N_1 \to \ell_i\, H} + \Gamma_{N_1 \to \bar\ell_i\, \bar H}}
\nonumber\\&
\simeq \frac{1}{8\, \pi}\, \frac{1}{(y_N^\dagger\, y_N)_{11}}\, \sum_{j=2, 3} \text{Im}\left(y_N^\dagger\, y_N\right)^2_{1j} \times \mathcal{F}\left(\frac{M_{N_j}^2}{M_{N_1}^2}\right),
\end{align}
with
\begin{equation}
    \mathcal{F}(x) \equiv \sqrt{x}\,\left[\frac{1}{1-x}+1-(1+x)\,\log\left(\frac{1+x}{x}\right)\right]\,.
\end{equation}
This can be further simplified to~\cite{Kaneta:2019yjn, Co:2022bgh}
\begin{equation}
    |\kappa_{\Delta L}| \simeq \frac{3\, \delta_\text{eff}}{16\, \pi}\, \frac{M_{N_1}\, m_{\nu_i}}{v^2}\simeq 
    10^{-6} \delta_{\rm eff}\left(\frac{M_{N_1}}{10^{10}}\right)\left(\frac{m_{\nu_i}}{0.05~\rm{eV}}\right)\,,    
\end{equation}
where $i=2\,,3$ for normal hierarchy and $\delta _{\rm eff}$ is the effective CP-violating phase (see Appendix.~\ref{sec:cp} for details) 

\begin{equation}
    \delta_\text{eff} = \frac{1}{(y_N)_{13}^2}\, \frac{\text{Im}(y_N^\dagger\,y_N)^2_{13}}{(y_N^\dagger\,y_N)_{11}}\,,  
\end{equation} 
whereas $v=174$ GeV is the Higgs vacuum expectation value. Note that a similar CP-asymmetry parameter can be obtained for the inverted hierarchy with $i=1\,,2$. We consider $m_{\nu_3}$ to be the heaviest active neutrino mass. The produced lepton asymmetry is eventually converted to baryon asymmetry via electroweak sphaleron processes, leading to baryon number yield at the point of evaporation \cite{Baumann:2007yr,Fujita:2014hha,Datta:2020bht}
\begin{align}
Y_B (\aev)=\frac{n_B}{s}\Big|_{\aev}=\mathcal{N}_{N_1}\,\kappa_{\Delta L}\,a_\text{sph}\,\frac{n_\text{BH} (\aev)}{s (\aev)}\,,
\label{eq:yB}
\end{align}
where $a_{\rm sph}=\frac{28}{79}$ and $n_{\rm BH} (\aev)$ is the PBH number density at the end of the evaporation process when the scale factor is $\aev$. 
\subsection{Leptogenesis during PBH reheating}
\label{sec:pbh-reheat}
\begin{figure}[htb!]
\includegraphics[scale=0.4]{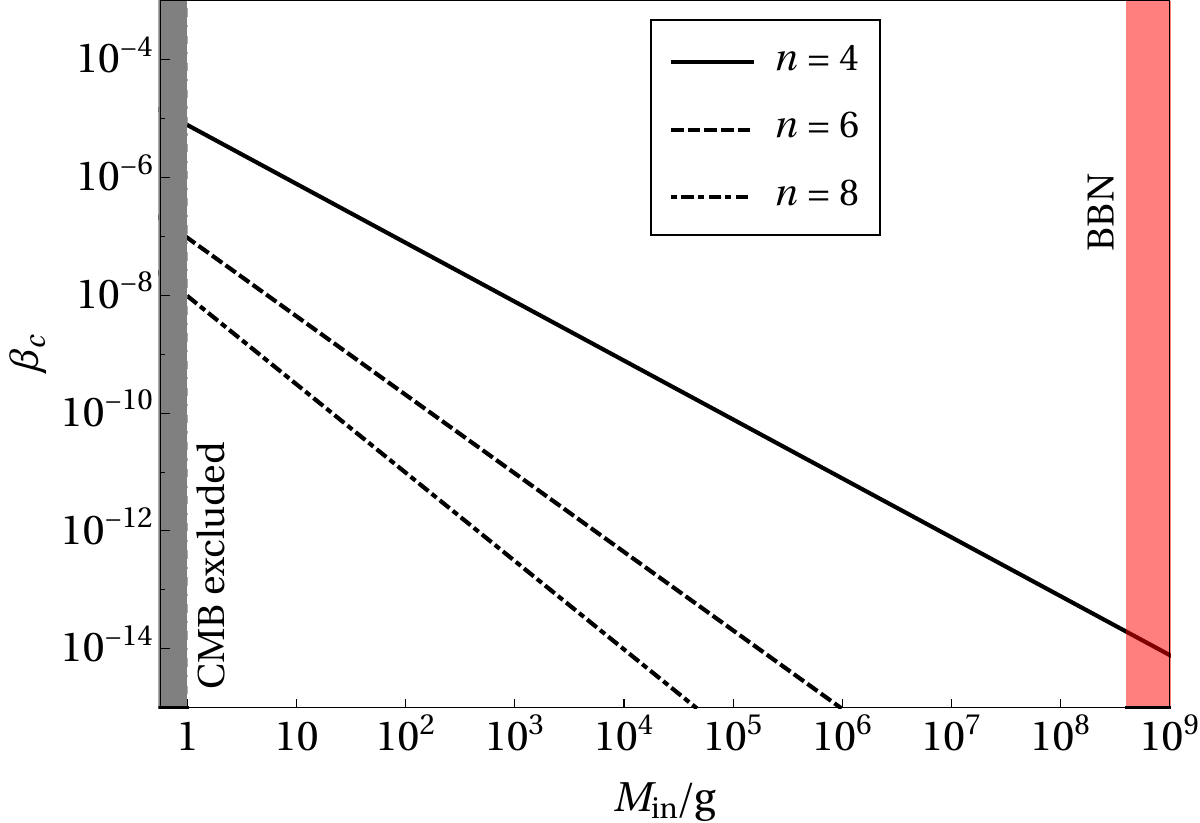}
\caption{$\beta_c$ as a function of $\Min$ for different choices of $n$, following Eq.~\eqref{Eq:betamin}, where the shaded regions are discarded from CMB bound on the inflationary Hubble scale (in gray) and BBN bound (in red).}
\label{fig:betac}
\end{figure}

In the PBH reheating scenario, the decay of PBHs is sufficient to reheat the Universe \cite{RiajulHaque:2023cqe}. If we assume no further entropy production after PBH evaporation,
the asymmetry is conserved and is given by $Y_B^0=Y_B(\aev)$.
However, depending on the initial population density, PBH reheating can be accomplished in two ways. If we define 
\begin{align}\label{eq:beta}
\beta=\frac{\rho_{\rm BH}}{\rho_{\rm tot}}\Big|_{\rm in}\,,    
\end{align}
as the ratio between PBH energy density and the background energy density at the point of formation, it was shown in~\cite{RiajulHaque:2023cqe,Haque:2024cdh} that in the presence of inflaton field
$\phi$, 
for $\beta$ larger than some critical value $\beta_{\rm c}$ given by
\begin{eqnarray}
\label{Eq:betamin}
\beta_c & = & \left[\frac{\epsilon}{2\pi\gamma(1+w_\phi)}\right]^{\frac{2w_\phi}{1+w_\phi}} \left(\frac{M_P}{\Min}\right)^{\frac{4w_\phi}{1+w_\phi}} \nonumber\\
&\simeq &\left(7.6\times 10^{-6}\right)^{\frac{4w_\phi}{1+w_\phi}}
\left(\frac{ 1\,\rm g}{\Min}\right)^{\frac{4w_\phi}{1+w_\phi}}\,,
\end{eqnarray}
where $w_\phi=(n-2)/(n+2)$ is the equation of state of an inflaton oscillating in a potential $V(\phi)\propto\phi^n$ (see Appendix~\ref{sec:inf} for details), PBH energy density dominates over that of 
inflaton {\it before} the evaporation process is complete. From Eq.~\eqref{Eq:betamin}, we find, $\beta_c \simeq 7.6\times 10^{-6}\left(\frac{1~\rm{g}}{\Min}\right)$ for $n=4$, 
and $\beta_c \simeq 9.3\times 10^{-8}\left(\frac{1~\rm{g}}{\Min}\right)^\frac43$ for $n=6$.  We show in Fig.~\ref{fig:betac} the variation of $\beta_c$ with $\Min$ for different choices of $n$.

On the other hand, if $\beta<\beta_{\rm c}$, even if
the reheating is mainly generated by the PBHs, 
the entire PBH evanescence takes 
place in an inflaton-dominated background. Indeed,
reheating can still be achieved through evaporating PBHs if the inflaton coupling 
is small enough and less than some critical value $y_{\rm \phi}^{\rm c}$ (See Appendix~\ref{sec:inf}). Additionally, it is also required that 
the inflaton  energy density redshifts faster than the radiation energy density, i.e., $w_\phi>1/3$ (or $n>4$).
This distinction between a PBH evaporation during (or not)
the inflaton domination era is crucial for the dilution of the different densities, and therefore for the baryonic asymmetry $Y_B(\aev)$.
\subsubsection{\bf Scenario-I: $\beta>\beta_{\rm c}$}
Solving the Boltzmann equation, we find for 
the evolution of inflaton density \cite{Garcia:2020eof,Garcia:2020wiy}
\begin{align}
& \rho_\phi=\rhoe\left(\frac{\aend}{a}\right)^{3(1+w_\phi)}=\rhoe\left(\frac{\aend}{a}\right)^{\frac{6n}{n+2}}\,,    
\end{align}
whereas
\beq
\rhobh=n_\BH M_\BH \propto a^{-3}\,.
\eeq 
This means that 
for $n>2$ and sufficiently long-lived PBH, the PBH can dominate the energy budget of the Universe 
at a time $t_{\rm BH}$ corresponding to the scale factor
$\abh$, before they evaporate. Since PBHs are formed during inflaton domination while they evaporate during PBH domination, from Eq.(\ref{Eq:mbht}), the PBH mass evolves as
\begin{align}\label{eq:abh}
&  M_{\rm BH}^3\simeq\Min^3-\frac{2\,\sqrt{3}\epsilon\,M_P^5}{\sqrt{\rho_\phi(\abh)}}\,\left(\frac{a}{\abh}\right)^{3/2}\,,  
\end{align}
where we assume $\aev \gg \abh$ for which $M_{\rm BH}(\abh)\simeq \Min$.
We then obtain the scale factor associated with the evaporation time ~\cite{RiajulHaque:2023cqe}
\beq \label{Eq:aevpbh}
\frac{\aev}{\abh}=\frac{\Min^2\rhoe^\frac{1}{3}}{(2 \sqrt{3}\epsilon M_P^5)^\frac{2}{3}}
\left(\frac{\ae}{\abh}\right)^{(1+w)}\,,
\eeq
imposing the condition $M(\aev)=0$, where $\abh$ corresponds to the onset of PBH domination~\cite{RiajulHaque:2023cqe}
\begin{equation}
\frac {a_{\rm BH}}{\aend}=\left(\frac{M_{\rm in}\,H_{\rm end}}{4\pi\gamma\,M_P^2}\right)^{\frac{2}{3(1+w_\phi)}}\beta^{-\frac{1}{3\,w_\phi}}\,.
\end{equation}
We will be exploiting these relations in Sec.~\ref{sec:grav-lepto} and Sec.~\ref{sec:PGW}.

For $\beta>\beta_{\rm c}$ evaporation is completed during PBH  domination; hence, reheating and evaporation points are identical, i.e., $T(\aev) \equiv T_{\rm RH}$. Since PBH domination behaves like a dust-like equation of state $(w_{\rm BH}=0)$, the Hubble parameter at the time of evaporation reads 
\begin{align}
& H(\aev)=\frac{2}{3\,t_{\rm ev}}=\frac{2\,\Gamma_{\rm BH}}{3}\,.   
\end{align}
Utilizing this expression, one can find the reheating temperature
\begin{align}
\frac{\rhorh}{3 M_P^2} = H^2(\aev)
\Rightarrow \trh=
\left(\frac{12\,\epsilon^2}{\alpha_T}\right)^{\frac{1}{4}}M_P\left(\frac{M_P}{\Min}\right)^{\frac{3}{2}}\,,
\label{Eq:trh1}
\end{align}
where $\alpha_T=\frac{\pi^2}{30}\,g_{\rm RH}$, and $g_{\rm RH}= 106.75$ is the degrees of freedom associated with the thermal bath at $\trh$. 

To determine the RHN yield at the point of evaporation, one needs to compute BH number density at $a_{\rm ev}$, which reads
\beq \label{Eq:nbhev}
n_{\rm BH} (\aev)\simeq\frac{\rho_\BH(\aev)}{\Min}=\frac{\rhorh}{\Min}=
\frac{12\,\epsilon^2M_P^{10}}{M_{\rm in}^7}\,.
\eeq
Eq.~\eqref{Eq:nbhev}, together with (\ref{Eq:trh1}) provides
\beq \label{Eq:ratioNT1}
\frac{n_\BH(\aev)}{T^3(\aev)}=\left(12\,\epsilon ^2\,\alpha_T^3\right)^{\frac{1}{4}}\left(\frac{M_P}{M_{\rm in}}\right)^{\frac{5}{2}}
\simeq 4\times 10^{-12}\left(\frac{1~\rm{g}}{\Min}\right)^\frac52\,.
\eeq
\begin{figure}[htb!]
\includegraphics[scale=0.42]{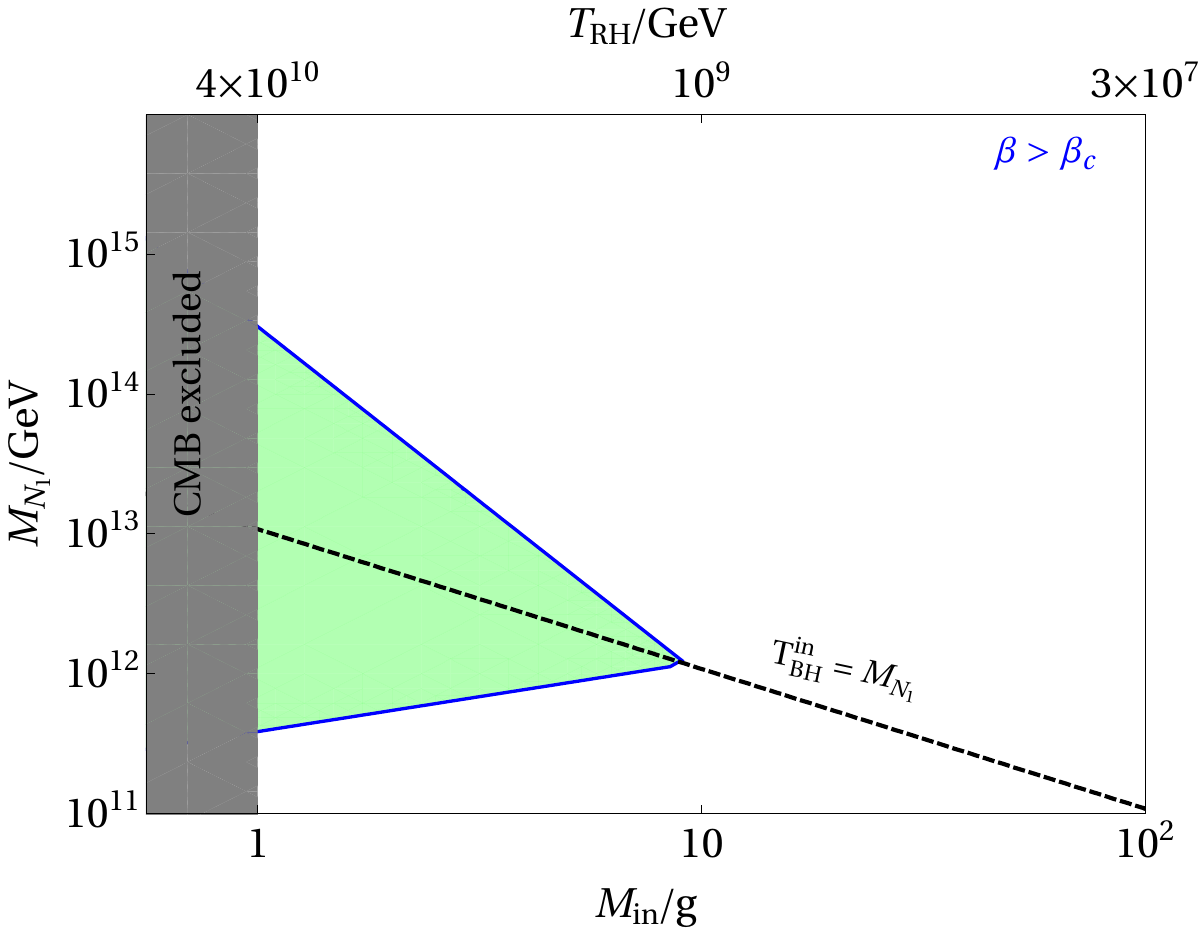}
\caption{Viable parameter space in the $[M_{N_1}\,,M_{\rm in}]$ plane, considering $\beta >\beta_c$. For each $\Min$ corresponding $\trh\,(\Tev)$ is mentioned along the top axis. The gray shaded region is disallowed from CMB bound on the scale of inflation for $n=6$ [cf. Eq.~\eqref{Eq:min}].}
\label{fig:fig1}
\end{figure}
One can then compute the final baryon  asymmetry utilizing Eq.~\eqref{eq:yB}
\begin{align}\label{eq:bgtbc}
& Y_B (T_{\rm 0})\simeq 8.7\times 10^{-11}\,\delta_{\rm eff}\,\left(\frac{m_{\rm \nu,\,max}}{0.05\,\rm eV}\right) 
\nonumber\\&
\begin{cases}
\left(\frac{M_{N_1}}{3.7\times 10^{11}\,\rm GeV}\right)\times\left(\frac{1\rm g}{M_{\rm in}}\right)^{\frac{1}{2}}\,,~~M_{N_1}<T_{\rm BH}^{\rm in}
\\
\left(\frac{3\times 10^{14}\,\rm GeV}{M_{N_1}}\right)\times\left(\frac{1\rm g}{\Min}\right)^{\frac{5}{2}}\,,~~M_{N_1}>T_{\rm BH}^{\rm in}\,,
\end{cases}
\end{align}
where $g_j=2$ for Majorana-like RHNs is considered.
 The interesting point is that,
for a given PBH mass, considering $\delta_{\rm eff} \lesssim \mathcal{O}(1)$, two regions are allowed for $M_{N_1}$. 

In Fig.~\ref{fig:fig1}, we show in green the 
allowed region of the parameter space ($\Min$,\,$M_{N_1}$) that can satisfy the observed baryon asymmetry for $\beta > \beta_c$. The slope 
of the two boundaries (in blue) is dictated by Eq.~\eqref{eq:bgtbc} setting $\delta_{\rm eff}=1$. 
We recognize the limit $M_{N_1}\propto \Min^{1/2}$ for $M_{N_1}<\Tbh^{\rm in}$, while for $M_{N_1}>\Tbh^{\rm in}$,  $M_{N_1}\propto \Min^{-5/2}$. Within the green shaded region, 
surrounded by the two boundaries, depending on the 
choice of $\delta_{\rm eff}$, 
it is possible to achieve the observed baryon asymmetry for a given $M_{N_1}$. The lower bound on PBH mass obtained 
from the gray-shaded region is set by the maximum energy scale of inflation, which is constrained by 
the CMB observation. It is important to note here
that the allowed region also satisfies the hierarchy $M_{N_1}>T(\aev)\equiv\trh$, validating non-thermal leptogenesis. 
Otherwise, for $M_{N_1}<T(\aev)$, the RHNs produced from PBH evaporation are in the thermal bath, and washout processes can not be neglected~\cite{Fujita:2014hha}. Thus, our first
result is that for $\beta>\beta_c$, the right 
baryon asymmetry is achievable for $10^{12}\lesssim M_{N_1}\lesssim 10^{15}$ GeV and 
PBH mass $\Min\lesssim\mathcal{O}(10)$ g. Heavier 
PBH masses do not have a sufficient {\it number} density to produce RHN in the right amount to fit with the measured $Y_B(T_0)$ as it is clear from Eq.(\ref{eq:bgtbc}).
\subsubsection{\bf Scenario-II: $\beta<\beta_c$}
For $\beta<\beta_{\rm c}$, PBHs are formed 
{\it and} evaporate during inflaton domination. 
In contrast to the previous case, PBHs never dominate the entire energy component of the 
Universe. Indeed, if the inflaton-matter 
coupling strength is less than some critical value $y_{\phi}^{\rm c}$ given in the appendix, Eq.(\ref{Eq:yphic}), and the inflaton equation of 
state mimics that of a  stiff fluid, $w_\phi>1/3$, 
there is a possibility for the PBHs to be responsible of the 
reheating even if not dominating 
the energy budget~\cite{RiajulHaque:2023cqe}. 
Another way of looking at things is to note that 
for a given Yuwaka coupling $y_{\phi}$, 
there always exists a threshold value of $\beta$, 
namely, $\beta_{\rm BH}$, above which PBH 
evaporation governs reheating temperature (see 
Appendix~\ref{sec:inf} for details). One can find 
the expression for $\beta_{\rm BH}$
for an inflaton potential $V(\phi)=\lambda M_P^4\left(\frac{\phi}{M_P}\right)^n$ as \cite{RiajulHaque:2023cqe}
\bea\label{eq:betaBH} 
&&
\beta_{\rm BH}=\left(\frac{y_{\rm \phi}\,\alpha_{n}}{8\pi}\right)^{\frac{2(3\,w_\phi-1)}{3(1-w_\phi)}}\,\left(\frac{48\pi^2}{\lambda}\right)^{\frac{1-3\,w_\phi}{3\,(1+w_\phi)}}
\\
&&
\left(\frac{\epsilon\,\gamma^{-3\,w_\phi}}{2\pi\,(1+w_\phi)}\right)^{\frac{2}{3\,(1+w_\phi)}}\left(\frac{M_P}{\Min}\right)^{\frac{2\,(1-w_\phi)}{(1+w_\phi)}}\,,
\nonumber
\eea
where $\bar{\alpha}_n=\frac{2\,(1+w_\phi)}{(5-9\,w_\phi)}\sqrt{\frac{6\,(1+w_\phi)\,(1+3w_\phi)}{(1-w_\phi)^2}}$.
The above equation is true for $n<7$, whereas for $n>7$, we have
\bea 
&&
\beta_{\rm BH}=-\frac{y_\phi^2\,\alpha_{n}}{8\,\pi}\,(48\pi^2)^{\frac{1-3\,w_\phi}{3\,(1+w_\phi)}}\,\lambda^{\frac{1-w_\phi}{2\,(1+\,w_\phi)}}
\\
&&
\left(\frac{\epsilon\,\gamma^{-3\,w_\phi}}{2\pi\,(1+w_\phi)}\right)^{\frac{2}{3\,(1+w_\phi)}}\left(\frac{M_P}{\Min}\right)^{\frac{2\,(1-w_\phi)}{1+w_\phi}}\left(\frac{\rho_{\rm end}}{M_P^4}\right)^{\frac{9\,w_\phi-5}{6\,(1+\,w_\phi)}}\,.
\nonumber 
\eea
\begin{figure*}[t!]
\includegraphics[scale=0.42]{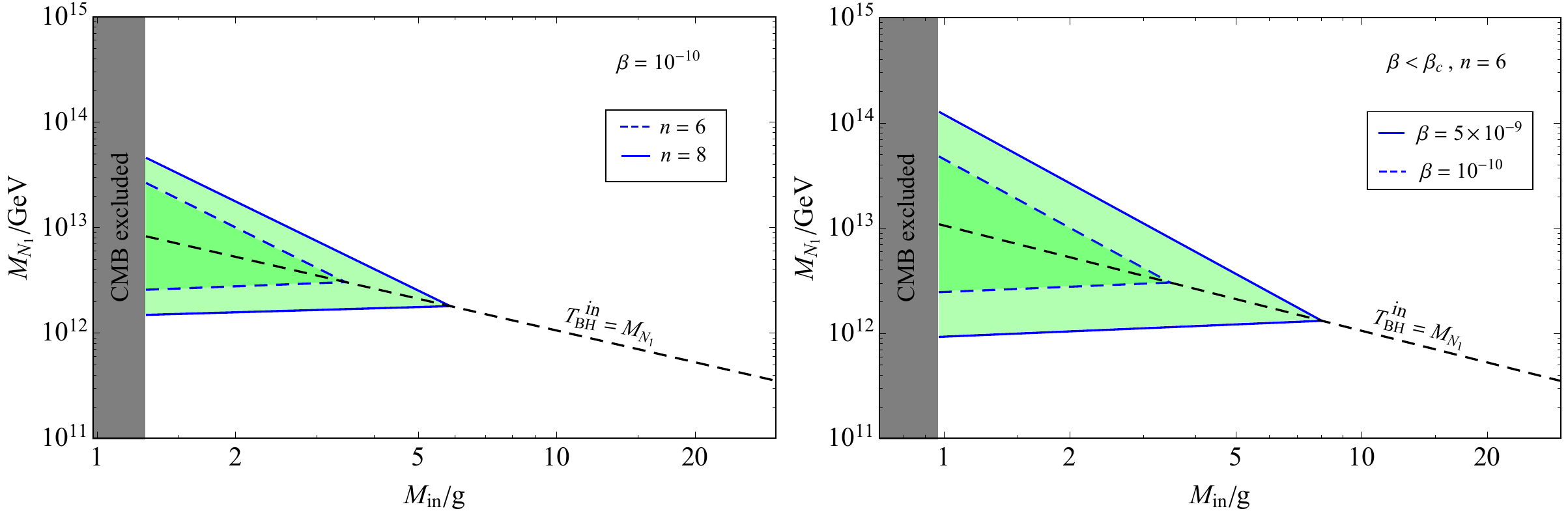}
\caption{{\it Left:} Allowed parameter space for $\beta=10^{-10}<\beta_c$, considering two different background equation of states, shown via different shades. {\it Right:} Same as left, but for a fixed $n=6$, considering two different $\beta$ values. In both plots, the gray shaded region is disallowed from CMB constraint on the scale of inflation [cf. Eq.~\eqref{Eq:min}].}
\label{fig:betalessbetac}
\end{figure*}
In this case, considering that after evaporation, all the PBH energy density is converted into the radiation energy density, $\rho_{\rm BH} (a_{\rm ev})\simeq\rho_{\rm R} (a_{\rm ev})=\alpha_TT_{\rm ev}^4$, we obtain
\begin{align}
& \frac{\nbh(\aev)}{\Tev^3} \simeq \left(\frac{\alpha_T^3\,\rbh(\aev)}{\Min^4}\right)^\frac{1}{4}\,. 
\label{Eq:rationbhT}
\end{align}
Since the PBHs behave like matter,
\beq \label{Eq:bhenergyden}
\rbh(\aev)=\beta \rho_\phi(\ain)\left(\frac{\ain}{\aev}\right)^3
=48 \pi^2 \gamma^2 \beta \frac{M_P^6}{\Min^2}\left(\frac{\ain}{\aev}\right)^3
\,,
\eeq
where we used from Eq.(\ref{Eq:min})
\beq
\rphi(\ain)=48\,\pi^2\,\gamma^2\,\frac{M_P^6}{\Min^2}\,.
\label{Eq:rhophiin}
\eeq
Considering that PBHs formed and evaporate during inflaton domination, we obtain
\begin{align}
\left(\frac{\ain}{\aev}\right)^3
=\left(\frac{H_{\rm ev}}{H_{\rm in}}\right)^{\frac{2}{1+w_\phi}}
=\left(\frac{\epsilon}{2\,(1+w_\phi)\,\pi\,\gamma}\,\frac{M_P^2}{\Min^2}\right)^\frac{2}{1+w_\phi}\,,
\label{Eq:ainaev}
\end{align}
where we used $H_{\rm in}$ given by 
Eq.(\ref{Eq:min}) and the Hubble parameter at evaporation
\begin{align}
H_{\rm ev}=H (a_{\rm ev})=\frac{2}{3(1+w_\phi)}\frac{1}{t_{\rm ev}}=\frac{2}{3(1+w_\phi)}\Gamma_{\rm BH}\,.    
\end{align}
Combining Eqs.~\eqref{Eq:bhenergyden} and~\eqref{Eq:ainaev}, we can compute the PBH energy density at evaporation time
\bea \label{Eq: radenergydensity}
&\rho_{\rm BH}(\aev)= 48\,\pi^2\,\beta\, \left(\frac{\gamma^{w_\phi}\times\epsilon}{2\pi\,(1+w_\phi)}\right)^\frac{2}{1+w_\phi}\,
M_P^4
\nonumber\\&\times
\left(\frac{M_P}{\Min}\right)^\frac{6+2w_\phi}{1+w_\phi}\,.
\label{Eq:pbhreheattemp}
\eea
Substituting the above expression into Eq.~\eqref{Eq:rationbhT}, we find
\beq\label{eq:ratioNT2}
\frac{n_{\rm BH}(\aev)}{T_{\rm ev}^3}=\left(48\,\pi^2\,\beta\,\alpha_T^3\right)^{\frac{1}{4}}\mu\,\left(\frac{M_{P}}{M_{\rm in}}\right)^{\frac{5+3\,w_\phi}{2(1+w_\phi)}}\,,
\eeq
where $\mu=\left(\frac{\gamma^{w_\phi}\,\epsilon}{2\,\pi(1+w_\phi)}\right)^{\frac{1}{2(1+w_\phi)}}$. Finally, using Eq.~\eqref{eq:yB}, we obtain
\begin{align}\label{eq:blsbc}
& Y_B (T_{\rm 0})\simeq 8.7\times 10^{-11}\,\delta_{\rm eff}\,\left(\frac{m_{\rm \nu,\,max}}{0.05\,\rm eV}\right)\mu \,\beta^{\frac{1}{4}}
\nonumber\\&
\begin{cases}
\left(\frac{M_{N_1}}{6.5\times10^8\,\rm GeV}\right)\times\left(\frac{M_P}{M_{\rm in}}\right)^{\frac{1-w_\phi}{2(1+w_\phi)}},M_{N_1}<T_{\rm BH}^{\rm in}
\\[8pt]
7\times10^{18}\,\left(\frac{6.5\times10^{8}\,\rm GeV}{M_{N_1}}\right)\times\left(\frac{M_P}{M_{\rm in}}\right)^{\frac{5+3\,w_\phi}{2\,(1+w_\phi)}}\,,~M_{N_1}>T_{\rm BH}^{\rm in}\,,
\end{cases}
\end{align} 
One important point is to note that, in contrast with the previous case, here, final asymmetry strictly depends on the $\beta$ value as well as the background equation of the state where PBHs are formed and evaporated. Whereas, for $\beta>\beta_{\rm c}$ case, baryon asymmetry only depends on the formation mass $M_{\rm in}$, 
and does not depend on the equation of state of the background. 

The viable parameter space corresponding to $\beta<\beta_c$ is shown in 
Fig.~\ref{fig:betalessbetac}, where, as before, 
the maximally allowed region satisfying the observed baryon asymmetry is shown in the $(\Min, M_{N_1})$ plane for $\beta=10^{-10}$ and different 
values of $n$ (left), and for $n=6$ and different values of $\beta$ (right). As in Fig.~\ref{fig:fig1}, the green shaded region represents the parameter space where the right baryonic asymmetry can be obtained by 
tuning $\delta_{\rm eff}$ accordingly. 
As expected, as the density of PBH is lower than the previous case ($\beta < \beta_c$), 
it is more difficult to generate a reasonable asymmetry $Y_B$. This results in a more restricted parameter space.
In the left panel, we see, a smaller equation of state (smaller $n$) restricts even more the parameter space for a given $\beta=10^{-10}$. 
Note that the slopes of the 
boundaries ($\delta_{\rm eff}=1$) follow Eq.~\eqref{eq:blsbc}, where, for $n=6$, $M_{N_1}\propto\Min^{1/6}$ when $M_{N_1}<\Tbh^{\rm in}$ and $M_{N_1}\propto\Min^{-13/6}$ for $M_{N_1}>\Tbh^{\rm in}$. For $n=8$,
we find $M_{N_1}\propto\Min^{1/8}$ for $M_{N_1}<\Tbh^{\rm in}$, and $M_{N_1}\propto\Min^{-17/8}$ in the other case. This comes from the fact that 
the inflaton field redshifted faster for a stiffer background (larger $n$). Consequently,
the relative density increment of PBHs is larger, which makes it easier to generate correct baryon asymmetry.
As a result, in order to obtain the right baryon asymmetry in a stiffer background, the RHN is required to be lighter for $M_{N_1}<\Tbh^{\rm in}$, while for $M_{N_1}>\Tbh^{\rm in}$, heavier RHN is needed since for a fixed $\Min$ we see
\begin{align}
& Y_B^0\propto \frac{n_{\rm BH}}{T^3}\Bigg|_{\rm ev}\times\frac{M_{N_1}\,m_{\nu,{\rm max}}}{v^2}
\begin{cases}
\left(\frac{\Min}{M_P}\right)^2\,,~M_{N_1}<T_{\rm BH}^{\rm in}\\[8pt]
\left(\frac{M_P}{M_{N_1}}\right)^2\,,~M_{N_1}<T_{\rm BH}^{\rm in}\,.
\end{cases}
\end{align}

On the other hand, as expected and clear from
Eq.~(\ref{eq:blsbc}), larger values of $\beta$ enlarge the possibility of obtaining the right $Y_B$.  In the right panel of 
Fig.~\ref{fig:betalessbetac}, we show the allowed 
parameter space with two different values of $\beta=\{5\times10^{-9},\,10^{-10}\}$, for a fixed $n=6$. 
The PBH yields at $\aev$ increases monotonically 
with $\beta$, see Eq.~\eqref{eq:ratioNT2}, hence,  
for a fixed $\Min$, a larger $\beta$ results in 
comparatively larger viable parameter space.

At this stage of our study, a preliminary conclusion would be that the necessary PBHs mass
, which allows for a viable baryogenesis through
their decay while still ensuring the reheating, is $\Min \lesssim 10$ g. Heavier PBHs 
are not sufficiently dense to generate the
observed $Y_B$ asymmetry. When the background is dominated by the inflaton field, the parameter space is even reduced due to a dilution affecting the PBH. We can now see how the parameter space can evolve if the reheating is
led by the inflaton decay.

\subsection{Leptogenesis during inflaton reheating}
We now analyze the scenario where the direct decay of inflaton takes the leading role in completing 
the reheating. We consider a minimal reheating process through the Yukawa interaction $y_\phi\,\phi\,\bar{f}f$ between the inflaton and 
the SM-like fermion fields\footnote{It is worth 
noting that such an interaction can lead to fermion preheating~\cite{Greene:1998nh}, for which 
the production of particles is resonantly suppressed, a consequence of Fermi-Dirac
statistics.}. For $\beta< \beta_c$,  one can 
obtain a critical value of the Yukawa coupling strength $y_\phi^c$, such that for 
$y_\phi>y_\phi^c$ the reheating is always 
determined by the inflaton decay (see Appendix.~\ref{sec:inf} for details). In this case, the parameter space will be even reduced, compared to the previous situations, due to the dilution effect between the evaporation end (when PBH decays) and the end of reheating (when the inflaton decays).

Indeed, if reheating occurs after PBH evaporation, then due to entropy injection between $\aev<a<\arh$, the final asymmetry reads 
\beq \label{eq:yB-reheat}
Y_B (T_{\rm 0})=Y_B (T_{\rm RH})=\mathcal{N}_{N_1}\,\epsilon_{\rm \Delta L}\,a_\text{sph}\,\frac{n_\text{BH} (\aev)}{s (\arh)}\left(\frac{a_{\rm ev}}{a_{\rm RH}}\right)^3\,.
\eeq
Connecting the scale factor from the point of evaporation to the end of reheating, one can find
\bea
&&
\frac{\aev}{\arh}=
\left(\frac{t_{\rm ev}}{t_{\rm RH}}\right)^\frac{2}{3(1+w_\phi)}
=\left[\frac{3(1+w_\phi)}{2}\frac{H_{\rm RH}}{\Gamma_{\rm BH}}\right]^\frac{2}{3(1+w_\phi)}
\nonumber
\\
&&
=\left(\frac{(1+w_\phi)}{2\sqrt{3}}\frac{\sqrt{\alpha_T}\,\trh^2}{M_P}\frac{\Min^3}{\epsilon\, M_P^4}\right)^\frac{2}{3(1+w_\phi)}\,,
\label{Eq:aevarh1}
\eea
where $H_{\rm RH}$ denotes the Hubble parameter at the end of reheating. Thus, the RHN number density at the end of reheating reads
\begin{align}
&\frac{n_N(\arh)}{\trh^3}=\mathcal{C}
\begin{cases}
\left( \frac{M_P}{T_{\rm RH}}\right)^{\frac{3w_\phi-1}{1+w_\phi}}\left( \frac{M_{\rm in}}{M_P}\right)^{\frac{1-w_\phi}{1+w_\phi}} \,,~M_{N_1}<T^{\rm in}_{\rm BH}
\\[8pt]
\frac{M_P^2}{M_{N_1}^2}\left( \frac{M_P}{T_{\rm RH}}\right)^{\frac{3w_\phi-1}{1+w_\phi}}\left(\frac{M_{P}}{M_{\rm in}}\right)^{\frac{1+3w_\phi}{1+w_\phi}},M_{N_1}>T^{\rm in}_{\rm BH}
\end{cases}
\end{align}
where $\mathcal{C}=\tilde \mu\frac{540\,g_j\,\zeta(3)\beta}{g_{\rm *}(T_{\rm BH})\pi^2}$, with $\tilde \mu = 
\left(\frac{\sqrt{\alpha_T}\,\gamma^{w_\phi}}{4\sqrt{3}\,\pi}\right)^\frac{2}{1+w_\phi}$. The final baryon asymmetry thus reads
\begin{align}\label{eq:YBinfreheat}
& Y_B (T_{\rm 0})\simeq 8.7\times 10^{-11}\,\delta_{\rm eff}\,\left(\frac{m_{\rm \nu,\,max}}{0.05\,\rm eV}\right)\left( \frac{M_P}{T_{\rm RH}}\right)^{\frac{3\,w_\phi-1}{1+w_\phi}}\tilde{\mu }\,\beta
\nonumber\\&
\begin{cases}
\left(\frac{M_{N_1}}{9.5\times10^7\,\rm GeV}\right)\times\left( \frac{M_{\rm in}}{M_P}\right)^{\frac{1-w_\phi}{1+w_\phi}}\,,~M_{N_1}<T_{\rm BH}^{\rm in}
\\[8pt] 6.2\times10^{13}\,\left(\frac{10^{15}\,\rm GeV}{M_{N_1}}\right)\times\left(\frac{M_P}{M_{\rm in}}\right)^{\frac{1+3\,w_\phi}{1+w_\phi}}\,,~M_{N_1}>T_{\rm BH}^{\rm in}\,.
\end{cases}
\end{align}
Note that, in the present framework, we are always interested in the case where PBHs are formed {\it during} the reheating, implying $\ain<\arh$.
They can evaporate before $(\aev<\arh)$ or after $(\aev>\arh)$ the end of reheating, it was shown in Ref.~\cite{RiajulHaque:2023cqe,Haque:2024cdh}, 
that both cases lead to the same result. This can be understood from the following argument. At the present epoch, the number density of any particle $j$, produced via PBH evaporation is given by
\begin{align}
& n_j(a_0)=n_j(a_{\rm ev})\,\left(\frac{a_{\rm ev}}{a_0}\right)^3=
n_{\rm BH}(a_{\rm ev})\,\mathcal{N}_j\,\left(\frac{a_{\rm ev}}{a_0}\right)^3\,,
\nonumber
\end{align}
where $n_j(a_{\rm ev})$ and $n_{\rm BH}(a_{\rm ev})$ are the particle and PBH number density at the point of evaporation. This can be further written as
\begin{align}
& n_j(a_0)
=n_{\rm BH}(a_{\rm in})\,\mathcal{N}_j
\left(\frac{a_{\rm in}}{a_0}\right)^3
\nonumber\\&
=n_{\rm BH}(a_{\rm in})\,\mathcal{N}_j\left(\frac{a_{
\rm in}}{a_{\rm RH}}\right)^3\left(\frac{a_{\rm RH}}{a_0}\right)^3
\,.
\end{align}
If the dilution is dominated by the same field (in this case, the inflaton) between $a_{\rm in}$ and $a_{\rm RH}$, the relic abundance does not depend on the epoch of evaporation. Thus, irrespective of $a_{\rm ev} < a_{\rm RH}$ or $a_{\rm ev} > a_{\rm RH}$, the final result shall remain unaffected. 
\begin{figure*}[t!]
\includegraphics[scale=0.42]{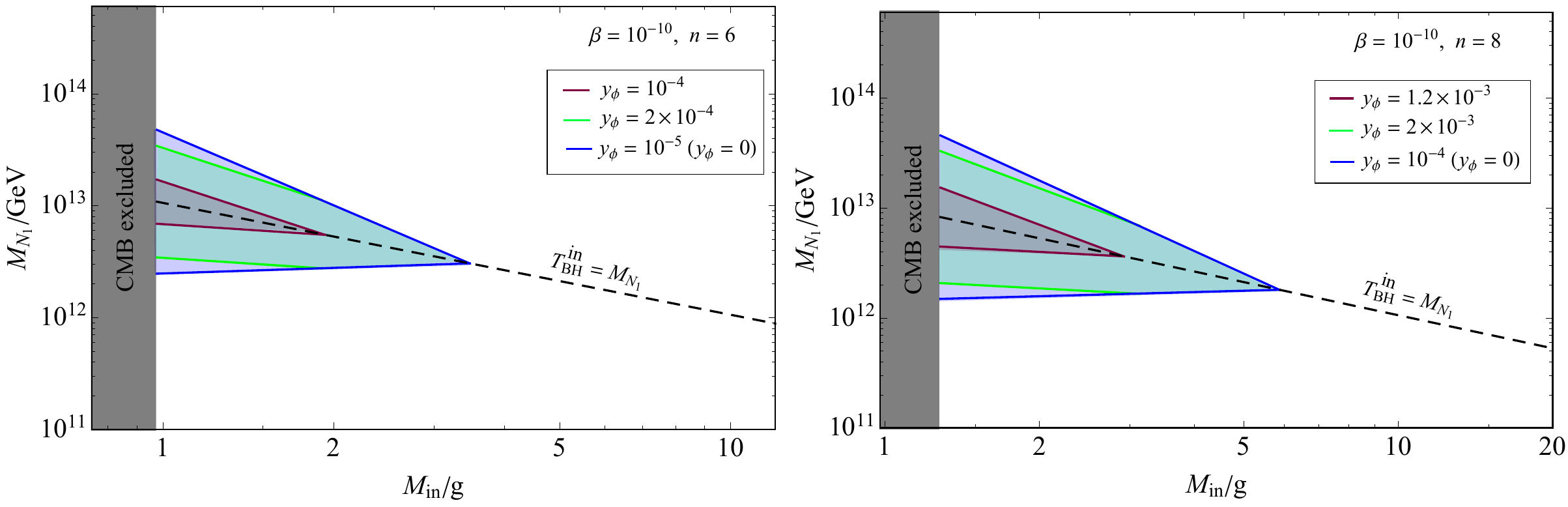}
\caption{{\it Left:} Parameter space showing observed baryon asymmetry for a fixed value of $n=6$, and three different choices of the Yukawa coupling $y_\phi$, shown via different shades. {\it Right:} Same as left, but for $n=8$. In both plots, the gray-shaded region is disallowed from CMB constraint on the scale of inflation [cf. Eq.~\eqref{Eq:min}]. For coupling strengths less than the critical value $y_\phi^c$ [cf. Eq.~\eqref{Eq:yphic}], below which PBH evaporation leads the reheating, behaves similarly as $y_\phi=0$. }
\label{fig:inflatonreheating}
\end{figure*}
\begin{figure}[htb!]
\includegraphics[scale=0.35]{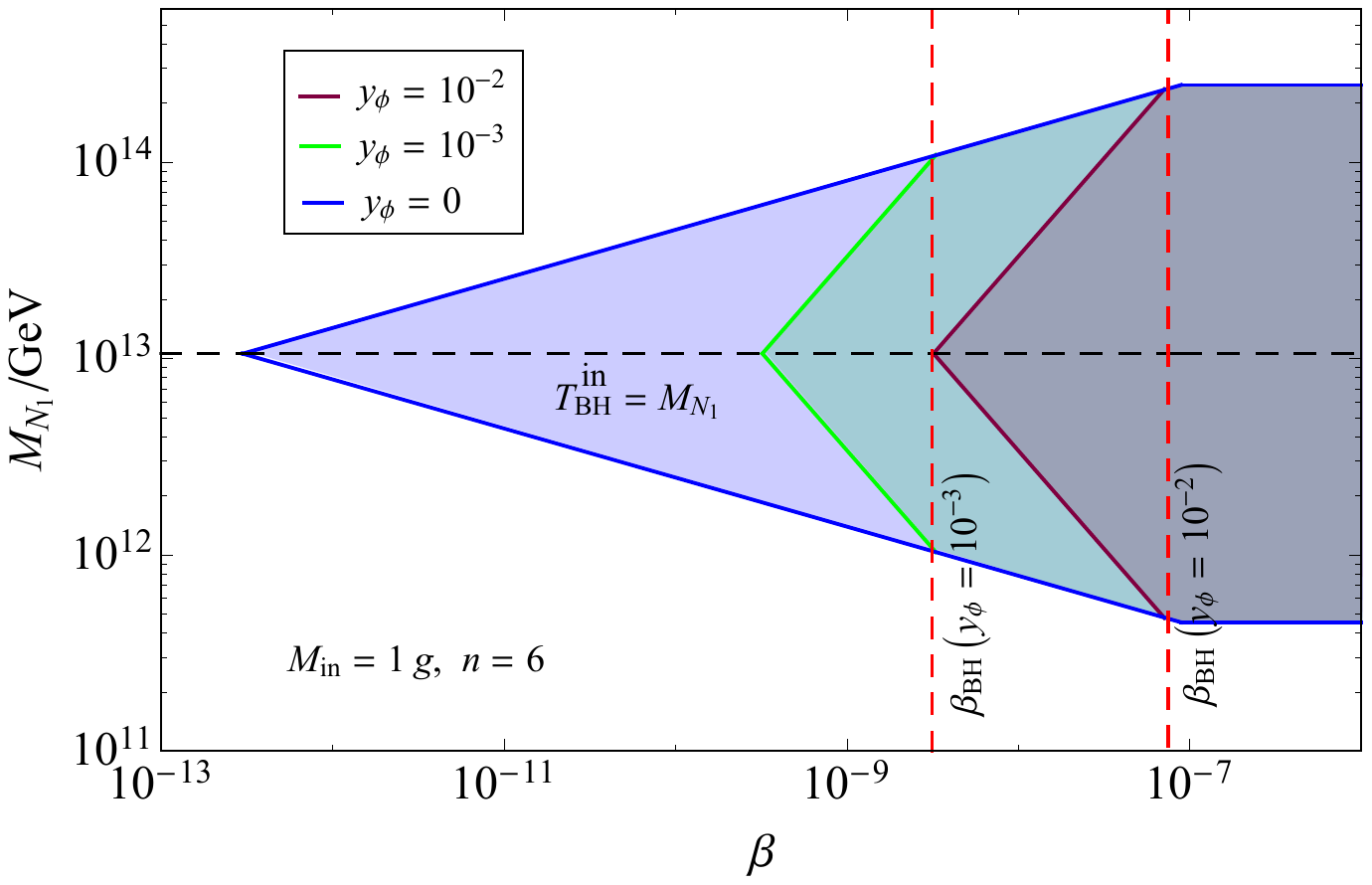}
\caption{Parameter space for right baryon asymmetry for a fixed BH mass $M_{\rm in}=1$ g and $n=6$, considering different choices of $y_\phi$, as shown via different shades. The red dashed line represents the threshold values of $\beta$, $\beta_{\rm BH}$ [cf. Eq.~\eqref{eq:betaBH}]  above which evaporating PBHs always dominate the reheating process. }
\label{fig:lep-beta}
\end{figure}

We show in Fig.~\ref{fig:inflatonreheating} 
the effect of inflaton reheating on the parameter 
space producing right baryon asymmetry in 
the $(\Min,M_{N_1})$ plane for fixed choices of $\beta$ and 
$y_\phi$, for $n=6$ (left) and $n=8$ (right). 
As expected, larger values of $y_\phi$ 
constrains even more the allowed parameter space. Indeed, a
stronger coupling results in higher $\trh$ due to the earlier decay of the inflaton, when it stores 
a larger amount of energy. This earlier decay tends 
to dilute even further the baryon asymmetry generated by the PBH decay. As a 
result, the right abundance is obtained with heavier 
RHNs for $M_{N_1}<\Tbh^{\rm in}$, and the opposite 
for $M_{N_1}>\Tbh^{\rm in}$. Increasing $n$ dilutes slightly more the inflaton before its decay, 
lowering $\trh$ and reopening very little parameter space, as one can see in the right panel of Fig.~\ref{fig:inflatonreheating} for $n=8$. It is interesting to note that since for $y_\phi<y_\phi^c$, PBH decay dominates over inflaton decay, hence all such couplings can effectively treated to be zero. This happens, for example, with $y_\phi=10^{-5}$ for $n=6$ and $\beta=10^{-10}$ as mentioned in the left panel of Fig.~\ref{fig:inflatonreheating}.

In Fig.~\ref{fig:lep-beta}, we show the viable parameter space for 
different choices of $y_\phi$ while fixing $\Min=1$ g and $n=6$. To the left of each vertical red 
dashed line, the Universe is reheated via 
inflaton. To the right we see, $M_{N_1}\propto\beta^{-1}$ for $M_{N_1}<\Tbh^{\rm in}$ and $M_{N_1}\propto\beta$ for $M_{N_1}>\Tbh^{\rm in}$, following 
Eq.~\eqref{eq:YBinfreheat}. Once $\beta>\beta_c$, the abundance depends only on the PBH lifetime and becomes independent of $\beta$. In conclusion, we showed
that if one wants to reconcile
a viable baryonic asymmetry generated by PBH decay, the mass spectrum should
lie in a region $M_\BH \lesssim 10$ g, whatever the reheating process. However, another minimal gravitational source of baryonic asymmetry exists, 
which is through the exchange of a graviton.

\section{Minimal Gravitational Leptogenesis in the presence of PBH}
\label{sec:grav-lepto}
\begin{figure}[htb!]
\includegraphics[scale=1.4]{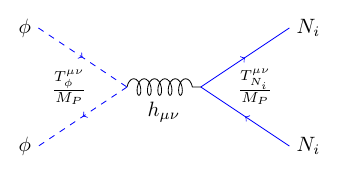}
\caption{Production of RHNs $N_i$ mediated by gravity during reheating, where $T^{\mu\nu}$ represents corresponding energy-momentum tensor.}
\label{fig:feyn}
\end{figure}

Another gravitationally sourced asymmetry  
involves the scattering of the inflaton during reheating, that leads to the production of the heavy RHNs through the exchange of a graviton, depicted in Fig.~\ref{fig:feyn}. 
This was studied in \cite{Co:2022bgh,Barman:2022qgt} and is considered as a
minimal, unavoidable source of leptogenesis.
In this section, we will not suppose any coupling
between the inflaton and the SM, except for the gravitational one ($y_\phi=0$).

Gravitational production can be achieved by considering the following interaction Lagrangian 
\beq
\sqrt{-g}\,\mathcal{L}_{\rm int}= -\frac{1}{M_P}\,h_{\mu \nu}\,\left(T^{\mu \nu}_{\rm SM}+T^{\mu \nu}_\phi + T^{\mu \nu}_{X} \right) \,,
\label{Eq:lagrangian}
\eeq
where $X$ is a particle that does not belong to the SM, which is a spin 1/2 Majorana fermion in the present context. The gravitational field can be realized by expanding the metric around Minkowski space-time $\eta_{\mu \nu}$ as $g_{\mu \nu}\simeq \eta_{\mu \nu}+\frac{2h_{\mu \nu}}{M_P}$, where $h_{\mu\nu}$ represents the canonically normalized quanta of the graviton. The graviton propagator for momentum $p$ is
\begin{equation}
 \Pi^{\mu\nu\rho\sigma}(p) = \frac{\eta^{\rho\nu}\eta^{\sigma\mu} + 
\eta^{\rho\mu}\eta^{\sigma\nu} - \eta^{\rho\sigma}\eta^{\mu\nu} }{2p^2} \, .
\end{equation}
The form of the stress-energy tensor $T^{\mu \nu}_i$ depends on the spin of the field and for Majorana spin-1/2 fermions $\chi$, takes the form
\begin{equation}\label{eq:tmunu}
T^{\mu \nu}_{1/2} =
\frac{i}{8}
\left[ \bar \chi \gamma^\mu \overset{\leftrightarrow}{\partial^\nu} \chi
+\bar \chi \gamma^\nu \overset{\leftrightarrow}{\partial^\mu} \chi \right] -g^{\mu \nu}\left[\frac{i}{4}
\bar \chi \gamma^\alpha \overset{\leftrightarrow}{\partial_\alpha} \chi
-\frac{m_\chi}{2}\,\overline{\chi^c} \chi\right]\,,
\end{equation}
whereas for a generic scalar $S$,
\begin{equation}
T^{\mu \nu}_{0} =
\partial^\mu S \partial^\nu S-
g^{\mu \nu}
\left[
\frac{1}{2}\partial^\alpha S\,\partial_\alpha S-V(S)\right]\,.
\label{eq:tmunuphi}
\end{equation}
As before, the heavy RHNs undergo CP-violating decay to produce the lepton asymmetry. Since we are considering non-thermal leptogenesis, hence we only take inflaton scatterings into account\footnote{This can be further ensured by noting that the thermalization rate $\Gamma^{\rm th}\simeq y_N^2\,T/(8\,\pi)$ remains below the Hubble rate during reheating at $T=M_1$. One can thus safely ignore the washout effects.}. 

For the production of $N_1$ through the scattering of the inflaton condensate, we consider the time-dependent oscillations of a classical inflaton field $\phi(t)$. The oscillating inflaton field with a time-dependent amplitude can be parametrized as
\begin{equation}
    \label{Eq:oscillation}
    \phi(t)= \phi_0(t)\cdot\mathcal{Q}(t) = \phi_0(t)\sum_{\nu=-\infty}^{\infty}\,{\cal Q}_n e^{-i\nu \omega t}\,,
\end{equation}
where $\phi_0(t)$ is the time-dependent amplitude that includes the effects of redshift and $\mathcal{Q}(t)=\sum_{\nu=-\infty}^{\infty}\,{\cal Q}_n e^{-i\nu \omega t}$ describes the periodicity of the oscillation of the inflaton field. The evolution of RHNs number densities $n_{N_i}$ is governed by the Boltzmann equation
\begin{align}\label{eq:dm-beq}
& \frac{dn_{N_i}}{dt} + 3\,H\,n_{N_i} = R^{\phi^n}_{N_i}\,,
\end{align}
where $R^{\phi^n}_{N_i}$ is the production rate of RHNs that we will mention in a moment. Defining the comoving number density as $Y_{N_i}=n_{N_i}\,a^3$, we can re-cast the Boltzmann equation as
\begin{align}\label{eq:dm-beq-comov}
& \frac{dY^T_{N_i}}{da}=\frac{a^2}{H}\,R^{\phi^n}_{N_i}\,.
\end{align}
The energy density of inflaton and radiation, on the other hand, evolves as
\begin{align}\label{eq:BErho}
&\frac{d\rho_\phi}{dt} + 3H\,(1+w_\phi)\, \rho_\phi = -(1+w_\phi)\,\Gamma_\phi\, \rho_\phi\,,
\nonumber\\
&\frac{d\rho_R}{dt} + 4H\, \rho_R = + (1+w_\phi)\,\Gamma_\phi\, \rho_\phi\,, 
\end{align}
where the production rate of radiation is given by~\cite{Clery:2021bwz, Clery:2022wib, Co:2022bgh}
\begin{align}
&   (1+w_\phi)\,\Gamma_\phi\, \rho_\phi = R^{\phi^n}_H \simeq \frac{N_h\rho_{\phi}^2}{16\pi M_P^4} \sum_{\nu=1}^{\infty}  2\nu\omega|{\mathcal{P}}^n_{2\nu}|^2 
\nonumber\\&
= \alpha_n M_P^5 \left(\frac{\rho_{\phi}}{M_P^4}\right)^{\frac{5n-2}{2n}} \, ,
   \label{Eq:ratephik}
\end{align}
where $N_h = 4$ is the number of internal degrees of freedom for one complex Higgs doublet, and we have 
neglected the Higgs boson mass. Here $\sum_{\nu=1}^{\infty}  2\nu\omega|{\mathcal{P}}^n_{2\nu}|^2 $ parametrizes the periodicity of oscillation of the inflaton potential. The values of $\alpha_n$ are computed following~\cite{Co:2022bgh,Barman:2022qgt}, and are given in Tab.~\ref{Tab:tablealphak}.
\begin{table}[htb!]\large
\centering
\begin{tabular}{c|| c}
\hline
$n$ & $\alpha_n$ \\
\hline\hline
6 &  0.000193 \\
\hline
8 & 0.000528 \\
\hline
10 & 0.000966 \\
\hline
12 & 0.00144\\
\hline
\end{tabular}
\caption{Relevant coefficients $\alpha_n$ for the gravitational leptogenesis [cf. Eq.~\eqref{Eq:ratephik}].}
\label{Tab:tablealphak}
\end{table}
Note that, to avoid conflict with the BBN that requires the reheating temperature $\trh\gtrsim 1$ MeV, one needs to 
consider\footnote{This requirement of having large $w_\phi$ can be relaxed with non-minimal 
gravitational couplings as discussed in~\cite{Co:2022bgh,Clery:2022wib,Barman:2022qgt}.} $w_\phi\gtrsim 0.65$~\cite{Haque:2022kez,Co:2022bgh} or $n\gtrsim 9$. However, it was shown in \cite{Clery:2021bwz}
that the gravitational wave constraints exclude a reheating with $\trh \gtrsim 2$ MeV with minimal 
gravitational coupling. It is then interesting to know if gravitational leptogenesis
is possible in the minimal case (pure exchange of a graviton) in the presence of PBH as the source of reheating. Both productions (from inflaton and from PBHs) being gravitational, 
they can be considered unavoidable sources of baryonic asymmetry in the early Universe. 
This should open 
a new window on the parameter space analyzed in 
the previous section.

\begin{figure*}[htp]
\centering
\subfigure[]{\includegraphics[scale=0.4]{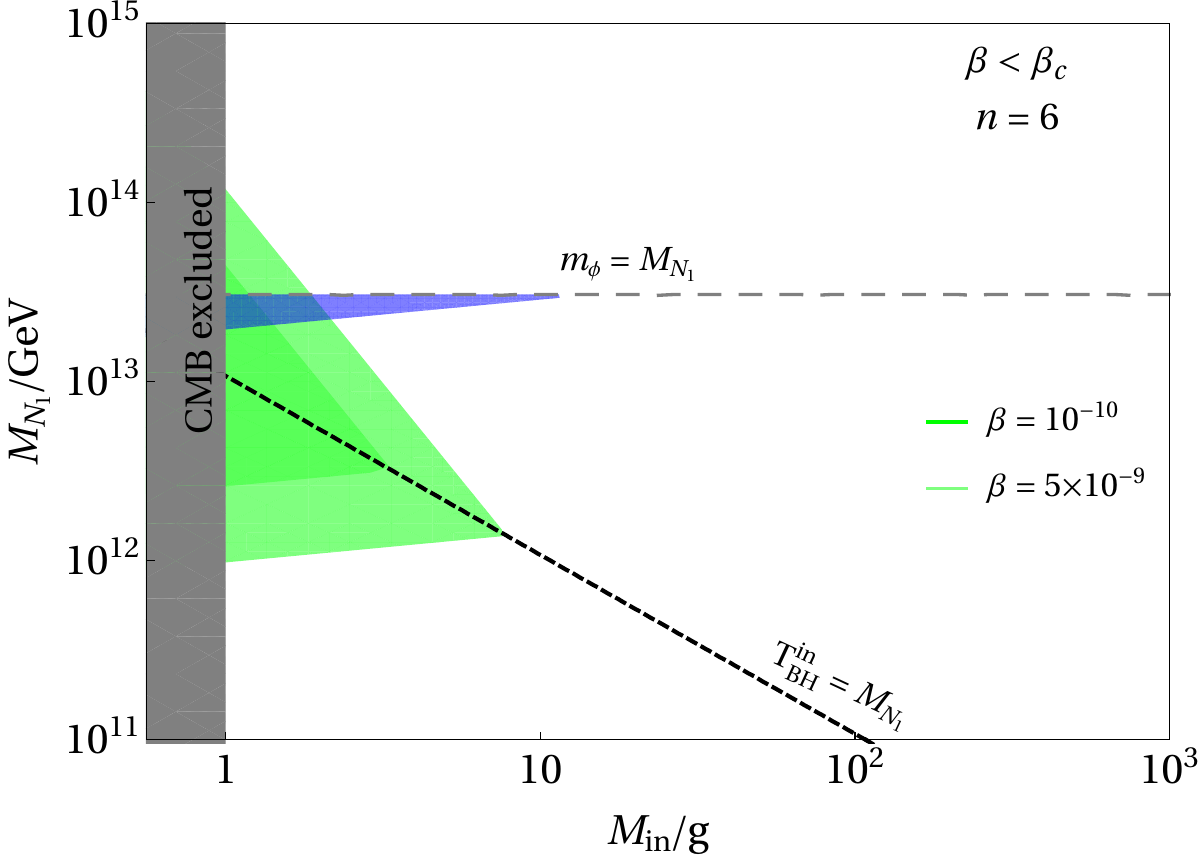}}
\subfigure[]{\includegraphics[scale=0.4]{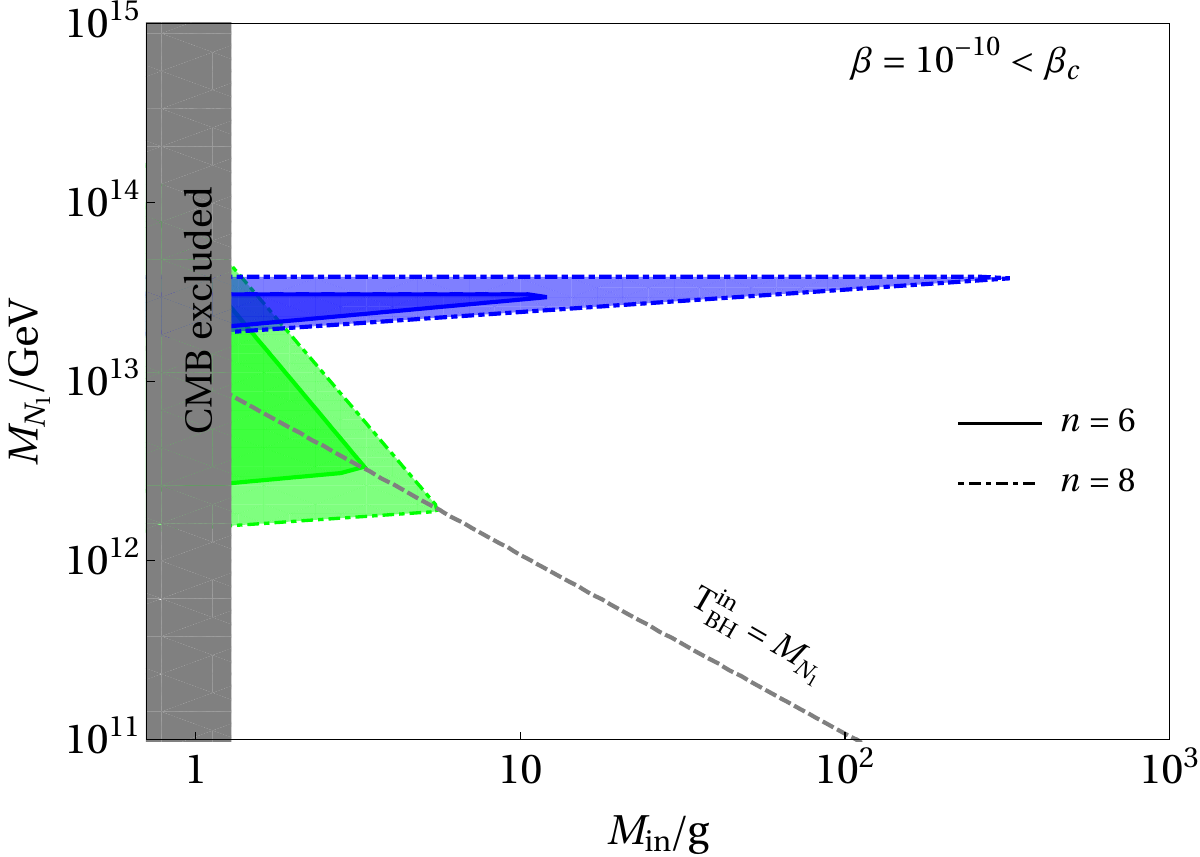}}
\caption{Inside the blue shaded region, the observed baryon asymmetry is obtained when {\it only} gravitational contribution from inflaton 
scattering is taken into account. In the {\it left} panel, we choose $n=6$ while considering two 
representative values of $\beta$. In the {\it right} panel, we choose $\beta=10^{-10}<\beta_c$
with two different values of $n=\{6,\,8\}$. In all cases, the green-shaded region corresponds to the 
viable parameter space for contribution from PBH evaporation alone, the gray-shaded region is 
forbidden from CMB bound on the scale of inflation [cf. Eq.~\eqref{Eq:min}], and we ensure reheating 
from PBH evaporation. The gray dashed line in the left panel indicates $m_\phi=M_{N_1}$, which is 
the kinematical limit for gravitational leptogenesis.
}
\label{fig:leptograv}
\end{figure*}

The production rate for $N_1$ from inflaton scattering mediated by gravity is given by~\cite{Clery:2021bwz} 
\begin{equation}
\label{eq:rateferm}
R^{\phi^n}_{N_1}=\frac{\rho_\phi^2}{4\pi M_P^4}\,\frac{M_{N_i}^2}{m_\phi^2}\,\Sigma_{N_i}^n \, ,
\end{equation}
where 
\begin{equation}
    \label{eq:ratefermion2}
    \Sigma_{N_i}^n = \sum_{\nu=1}^{+\infty} |{\cal P}_{2\nu}^n|^2\,
    \frac{m_\phi^2}{E_{2n}^2}\,
\left(1-\frac{4\,M_{N_1}^2}{E_{2\nu}^2}\right)^{3/2} \,,
\end{equation}
accounts for the sum over the Fourier modes of the inflaton potential, and $E_\nu = \nu \omega$ is the energy of the $n^{\rm th}$ inflaton oscillation mode. The full expression for the inflaton mass $m_\phi$ can be found in Appendix~\ref{sec:inf}.

Since we are only concerned about $N_1$ production, the comoving number density of $N_1$ during the post-inflationary era is given by
\begin{equation}\label{eq:RHN-comovingnumber}
\frac{dY^{\phi^n}_{N_1}}{da}=a_{\rm end}^2\frac{\sqrt{3}\,M_{N_1}^2\,M_P}{4\pi\, n\,(n-1)\,\lambda^{\frac{2}{n}}}\left(\frac{\rho_{\rm end}}{M_P^4}\right)^{\frac{n+4}{2\,n}}\left(\frac{a}{a_{\rm end}}\right)^{-\frac{n+8}{n+2}}\,\Sigma_{N_1}^n\,,
\end{equation}
where we have considered the fact that the Hubble expansion has the dominant contribution from inflaton energy density during reheating. 
Integrating Eq.~\eqref{eq:RHN-comovingnumber} between $a_{\rm end}$ and $a$ leads to RHN number density as
\begin{equation}
n_{N_1}^{\phi^n}(a)\simeq\frac{M_{N_1}^2M_P\sqrt{3}\,(n+2)}{24\pi n(n-1)\lambda^{\frac{2}{n}}}\left(\frac{\rho_{\rm end}}{M_P^4}\right)^{\frac{n+4}{2\,n}}\left(\frac{a}{a_{\rm end}}\right)^{-3}\Sigma_{N_1}^n\,,
\end{equation}
for $a\gg\aend$. 

For $\beta<\beta_{\rm c}$, as the inflaton energy density dominates for $\aend\ll a<\arh$, we 
then obtain
\begin{equation}
n_{N_1}^{\phi^n}(a_{\rm RH})\Big|_{\beta<\beta_c}\simeq \frac{M_{N_1}^2\,\sqrt{3}\,(n+2)\,\rho_{\rm RH}^{\frac{1}{2}+\frac{2}{n}}}{24\,\pi\,n(n-1)\lambda^{\frac{2}{n}}\,M_P^{1+\frac{8}{n}}}
\left(\frac{\rho_{\rm end}}{\rho_{\rm RH}}\right)^{\frac{1}{n}}\,
\Sigma_{N_1}^n\,.
\label{Eq:num-den-inf}
\end{equation}
On the other hand, for $\beta>\beta_{\rm c}$, there is an intermediate PBH-dominated phase before evaporation (reheating) that leads to

\begin{align}
& n_{N_1}^{\phi^n}(a_{\rm RH})\Big|_{\beta>\beta_c}\simeq \frac{M_{N_1}^2\,M_P\,(n+2)\,48^{\frac{1}{n}}}{8\,\pi\,n(n-1)\lambda^{\frac{2}{n}}\,\beta}\,\left(\frac{\rho_{\rm end}}{M_P^4}\right)^{\frac{1}{n}}
\nonumber\\&
\left(\frac{M_P}{\Min}\right)^\frac{2+5n}{n}\,\epsilon^2\,(\pi\gamma)^{-1+\frac{2}{n}}\,\Sigma_{N_1}^n\,.
\end{align}
Note that for PBH domination, the number density has explicit $\beta$ dependence, which is expected since $\beta$ controls the PBH-dominated phase. The final asymmetry in the case of minimal gravitational leptogenesis thus becomes 
\begin{equation}\label{eq:yld-grav}
 \frac{Y_B(T_0)}{8.7\times 10^{-11}}\simeq \delta_{\rm eff}\left(\frac{m_{\nu,\text{max}}}{0.05 \,\text{eV}}\right)\frac{M_{N_1}}{1.1\times 10^8 \text{GeV}}\frac{n_{N_1}^{\phi^n}(a_{\rm RH})}{T_{\rm RH}^3}\,,
\end{equation}
where for $\beta<\beta_{\rm c}$,
\begin{align}\label{eq:blsbcgrav}
&\frac{n_{N_1}^{\phi^n}(a_{\rm RH})}{T_{\rm RH}^3}=\frac{\alpha_T^{\frac{n+2}{2n}}\sqrt{3}(n+2)}{24\,\pi\,n\,(n-1)\lambda^{\frac{2}{n}}}\left(\frac{M_{N_1}}{M_P}\right)^2\left(\frac{T_{\rm RH}}{M_P}\right)^{\frac{4-n}{n}}
\nonumber\\&
\times \left(\frac{\rho_{\rm end}}{M_P^4}\right)^{\frac{1}{n}}\Sigma_{N_1}^n\,.
\end{align}
To compute $\trh$ appearing in Eq.(\ref{eq:blsbcgrav}), one needs to 
compute the density of energy when the radiation produced by the PBH decay 
at $\aev$ dominates over the inflaton density.
In other words, we need to solve $\rhorh=\rhobh(\aev)\left(\frac{\aev}{\arh}\right)^4=\rho_\phi(\ai)\left(\frac{\ai}{\arh}\right)^{3(1+w_\phi)}$.
We obtain
\beq
\trh \propto\beta^\frac{3\,(1+w_\phi)}{3\,w_\phi-1}\left(\frac{M_P}{\Min}\right)^{\frac{3(1-w_\phi)}{2(1-3w_\phi)}}\,.
\label{Eq:trhapprox}
\eeq
The details of the calculation
is reported in Appendix.~\ref{sec:inf} [cf. Eq.~\eqref{Eq:pbhreheattempbis}]. Note that Eq. \eqref{eq:yld-grav}, \eqref{eq:blsbcgrav} is also true for reheating happening entirely from inflaton, where a particular coupling $y_\phi$ between the inflaton and the SM particles determines the reheating temperature [cf. Eq.~\eqref{Eq: reheatinflaton}]. 

For $\beta>\beta_{\rm c}$, we find
\begin{align}\label{eq:bgtbcgrav}
& \frac{n_{N_1}^{\phi^n}(a_{\rm RH})}{T_{\rm RH}^3}= \frac{48^{\frac{1}{n}}\,\epsilon^{\frac{1}{2}}\,\alpha_T^{\frac{3}{4}}\,(n+2)}{16\sqrt{2}\times3^{\frac{3}{4}}\pi\,n(n-1)\lambda^{\frac{2}{n}}\,\beta}\,\left(\frac{\rho_{\rm end}}{M_P^4}\right)^{\frac{1}{n}}
\nonumber\\&
\left(\frac{M_{N_1}}{M_P}\right)^2\left(\frac{M_P}{\Min}\right)^\frac{n+4}{2n}\,(\pi\gamma)^{-1+\frac{2}{n}}\,\Sigma_{N_1}^n\,.
\end{align}
Note that, for $\beta<\beta_c$, the PBH reheating scenario is only valid for $n>4$, as in this case, a faster dilution of the inflaton energy density is required compared to radiation to achieve successful reheating. This is also reflected in Eq.~\eqref{Eq:pbhreheattempbis}.

In Fig.~\ref{fig:leptograv}, on top of leptogenesis from PBH evaporation (in green), we show the contribution from minimal
gravitational leptogenesis in blue for different
values of $\beta<\beta_c$ (left) and different value of $n$ (right). 
This plot includes the dominant gravitational sources of baryonic asymmetry 
in a universe populated by an inflaton field and PBHs. For $\beta>\beta_c$, we find that inflaton gravitational production starts contributing on top of PBH 
evaporation for very light PBHs, which  
are in tension with the CMB bound. This can be understood from the fact that as lighter PBHs 
decay earlier, they cause less entropy dilution to the asymmetry compared to heavier PBHs that have 
longer lifetime. Therefore, for $\beta>\beta_c$, gravitational leptogenesis is important when the 
PBH mass is typically $\lesssim 1$ g. We, therefore 
refrain from showing the resulting parameter space for $\beta>\beta_c$ in Fig.~\ref{fig:leptograv}. 

For $\beta<\beta_c$, 
we see that minimal gravitational leptogenesis 
from inflaton scattering is more important for lighter PBHs. 
Combining Eqs.(\ref{eq:yld-grav}) and (\ref{Eq:trhapprox}),
we can write $Y_B^0\propto M_N^3\times \Min^{-\frac{3}{n}}$, which corresponds to $M_N\propto \Min^{-3/8}\,M_N^3$ for $n=8$, which is what we 
effectively observe in the figure. 
This comes from the fact that when inflaton dominates the energy budget of the Universe, the reheating temperature due to PBH evaporation evolves as
$\Min^{\frac{3}{n-4}}$, which one can see from Eq.(\ref{Eq:trhapprox}). 
Following Eq.~\eqref{eq:blsbcgrav}, a lighter $\Min$ implies lower reheating temperature, leading to larger yields. For the dependence on $\beta$, we note that $Y_B^0\propto \Min^{-3/n}\,\beta^{-3/4}$, using 
Eq.~\eqref{eq:blsbcgrav}, together with Eq.~\eqref{Eq:trhapprox}. Therefore, for a given $\beta$, gravity-mediated leptogenesis becomes significant
for lighter PBHs when $\beta<\beta_c$.
While the inflaton generates a sufficient amount of asymmetry, the PBH ensures a viable reheating if $y_\phi<y_\phi^c$. 
In every case, the RHN mass is restricted
to lie in the range
$5 \times 10^{11}~\text{GeV} \lesssim M_{N_1} \lesssim 10^{14}~\rm{GeV}$ when taking into account both (PBH and inflaton) contributions. Notably, gravitational leptogenesis is kinematically viable only when $m_\phi>M_{N_1}$, as denoted by the gray dashed line in the left panel.
\section{Primordial Gravitational Wave from Inflation}
\label{sec:PGW}
Gravitational waves are transverse ($\partial_ih_{ij} = 0$) and traceless ($h_{ii} = 0$) metric perturbations $ds^2 = a^2(t)\,(dt^2-(\delta_{ij}+h_{ij})\,dx^i\,dx^j)$. Their energy density spectrum (at sub-horizon scales) is defined as~\cite{Boyle:2005se,Watanabe:2006qe,Saikawa:2018rcs,Caprini:2018mtu}
\begin{align}
\Omega_{\text{GW}}(t, k) \equiv\dfrac{1}{\rho_{\text{crit}}}\dfrac{d\rho_{\text{GW}}(t,k)}{d\ln k} =\frac{k^2}{12a^2(t)H^2(t)}\Delta_h^2(t,k)\,,  \label{eq:GWenergySpectrum}
\end{align}
where $\Delta_h^2(t,k)$ is the tensor power spectrum at arbitrary times, defined by
\begin{align}
\left<h_{ij}(t,\mathbf{x})h^{ij}(t,\mathbf{x})\right>&\equiv\int \frac{dk}{k} \Delta_h^2(t,k)\,,
\end{align}
with $\left<...\right>$ denoting an average over a statistical ensemble. One can factorize the tensor power spectrum as~\cite{Caprini:2018mtu}
\begin{align}\label{eq:transferfunc}
\Delta_h^2(t,k)\equiv T_h(t,k) \Delta_{h,\text{inf}}^2(k)\,, 
\end{align}
with  $T_h(t,k)$ being the transfer function 
\begin{align}
T_h(t,k)=\frac{1}{2}\,\left(\frac{a_{\rm hc}}{a}\right)^2\,,  
\end{align}
where the factor of 1/2 appears due to the average over the tensor fluctuations. Here ``hc"  indicates the epoch of horizon re-entry (crossing) of a particular mode, and $\Delta_{h,\text{inf}}^2(k)$ represents the primordial tensor power spectrum from inflation~\cite{Boyle:2005se,Caprini:2018mtu}
\begin{align}
\Delta_{h,\text{inf}}^2(k) \simeq {\frac{2}{\pi^2}}\left(\frac{H_{\rm end}}{M_P}\right)^2\left(\frac{k}{k_p}\right)^{n_t}\,,
\label{eq:InfSpectrum}
\end{align}
with $n_t$ a spectral tilt, $k_{p}$ a pivot scale of the order the Hubble rate at the time of CMB decoupling, and $H_{\rm end}$ the Hubble rate when the mode $k_p$ exited the Hubble radius during inflation. Since we assume de-sitter-like inflation, the Hubble parameter is the same throughout inflation, and the spectral tilt turns out as $n_t\simeq0$. Note that such an assumption works fine with any slow-roll model of inflation, such as the $\alpha-$ attractor model of inflation. To determine the energy scale at the end of the inflation, we assume the $\alpha-$ attractor model of inflation as a sample model (for the form of the potential, see Eq. (\ref{eq: attractor}) in Appendix-\ref{sec:inf}). Most of the slow-roll model of inflation (including the one we are considering) behaves as $\phi^{2n}$ at the minima, and the average inflaton equation of state can be written as $w_\phi=(n-1)/(n+1)$. More specifically, the value of $w_\phi$ only depends on the behavior of the inflationary potential at the minima, which is related to the post-inflationary behavior of the potential. Hence, it can not capture any details of the inflationary model on the large scales.\\ Let us assume for a moment that, immediately after inflation, the Universe became radiation-dominated. The resulting GW energy density spectrum at the present epoch would then be scale-invariant for the frequency range corresponding to the modes crossing the Hubble radius during RD. However, if prior to RD, there is a non-standard phase, say reheating, the resulting present-day GW energy density spectrum consists of two parts: a tilted branch, corresponding to the modes that crossed the horizon {\it during} reheating, and a scale-invariant branch corresponding to the modes that crossed the horizon during RD. The spectral tilt of the PGW spectrum takes the form $ n_{\rm GW} =\frac{6\,w_\phi-2}{1+3\,w_\phi}$~\cite {Mishra:2021wkm,Haque:2021dha} that predicts a red-tilted spectrum for an EoS $w_\phi<1/3$, blue-tilted for $w_\phi>1/3$ and a scale-invariant spectrum for $w_\phi=1/3$. 
\begin{figure*}[htp]
\centering
\subfigure[]{\includegraphics[scale=0.082]{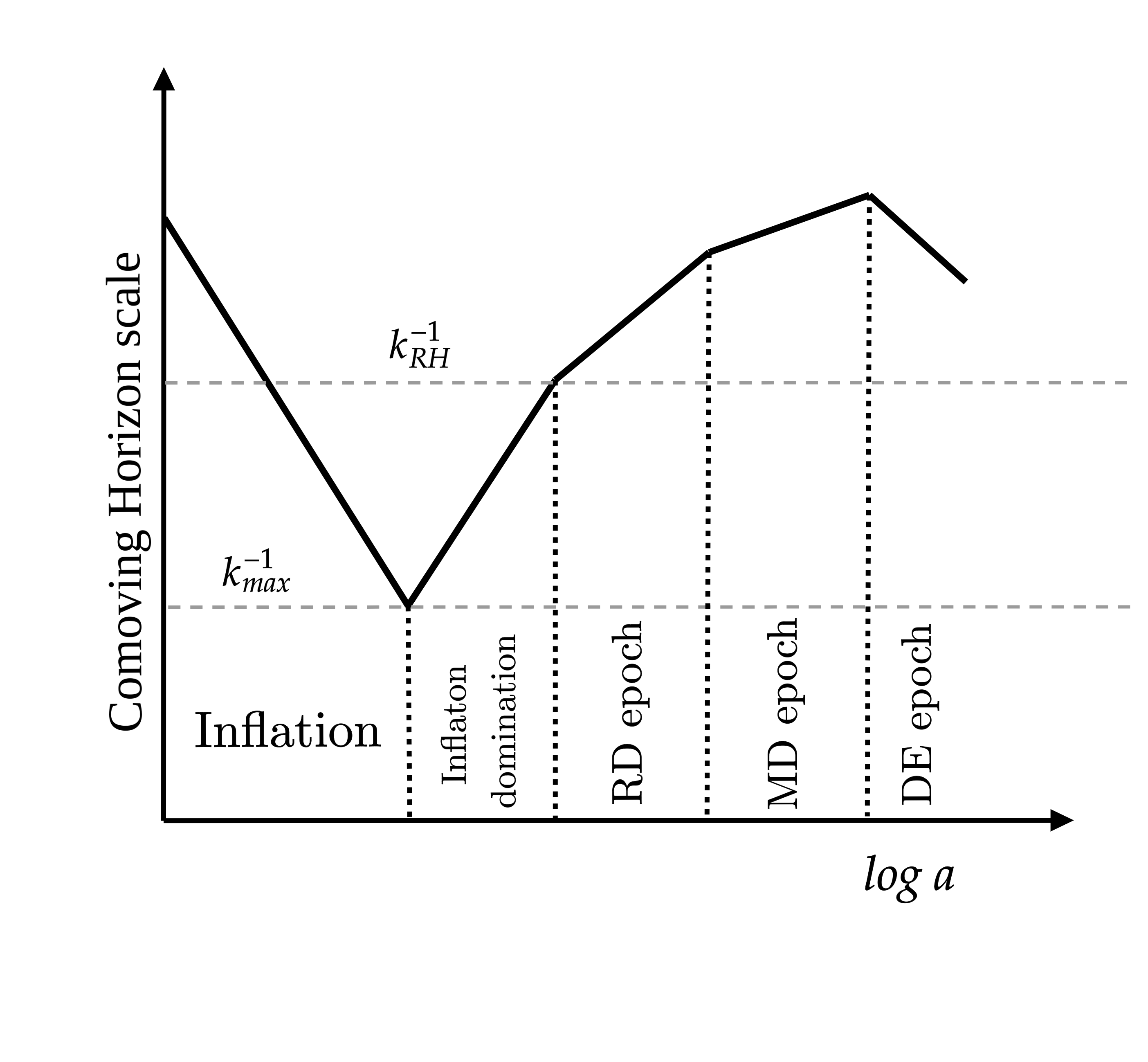}}
\subfigure[]{\includegraphics[scale=0.082]{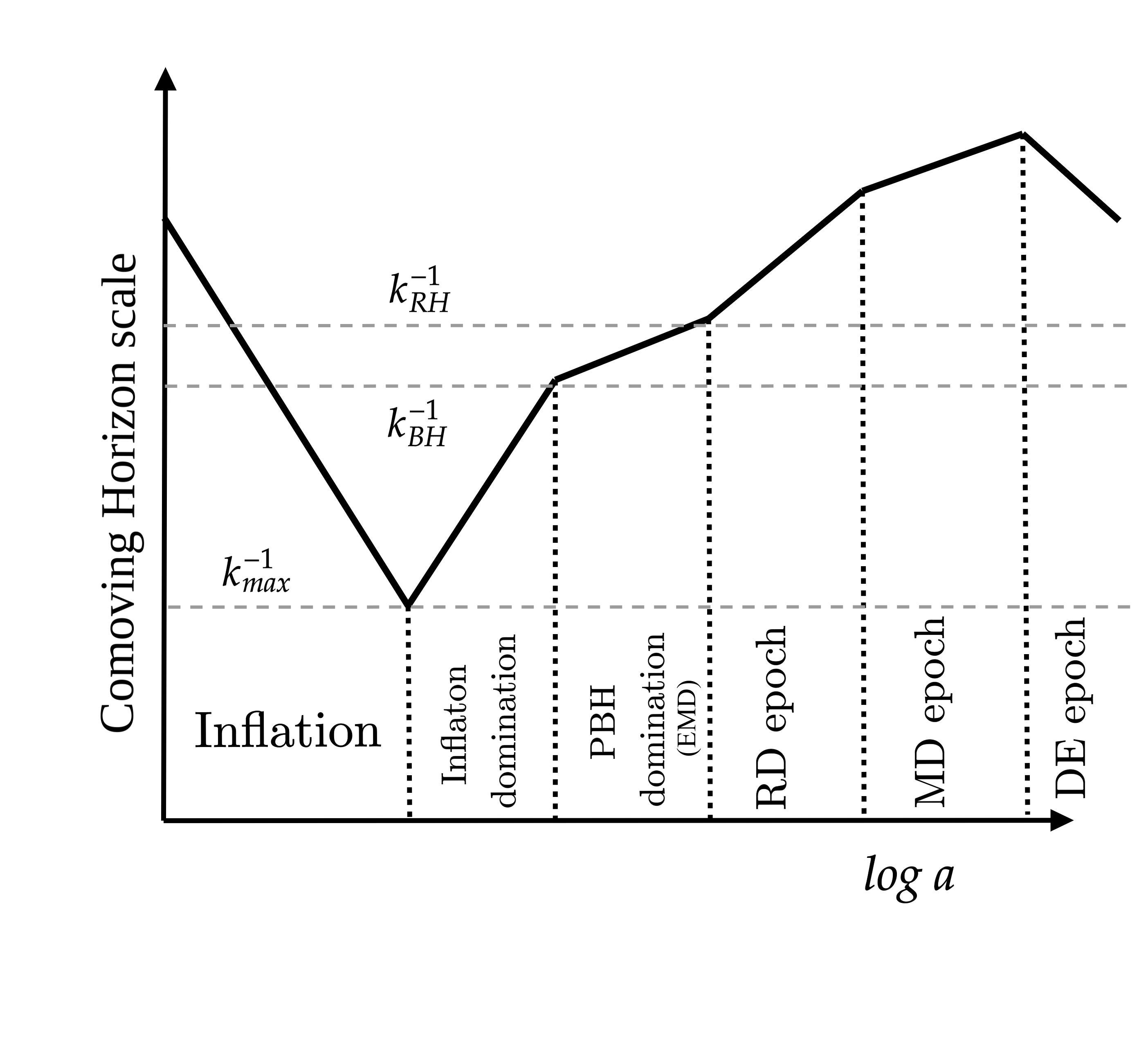}}
\caption{Schematic diagram showing the evolution of the comoving horizon scale $1/aH$ from inflation till today with respect to the scale factor. In the left panel, we consider $\beta<\beta_c$, while in the right, $\beta>\beta_c$. Here ``RD", ``MD, and ``DE" stand respectively for standard radiation domination, late matter domination, and dark energy. In the right panel, ``EMD" in parenthesis stands for early matter domination, corresponding to the PBH domination epoch. We also denote momenta corresponding to different epochs via the gray dashed horizontal lines.}
\label{fig:scheme}
\end{figure*}

In our present analysis, depending on $\beta$ value, we have two different scenarios as depicted in Fig.~\ref{fig:scheme}:
\begin{itemize}
\item For $\beta<\beta_c$, the Universe does not go through any PBH domination, 
\begin{align*}
\textit{Inflation}\to\textit{Reheating}\to\textit{Radiation domination}\,. 
\end{align*} 
\item For $\beta>\beta_c$, on the other hand, we have
\begin{align*}
&\textit{Inflation}\to\textit{Inflaton domination}\\&\to\textit{PBH domination}\to\textit{Radiation domination}\,. 
\end{align*}
\end{itemize}
In case when there is {\it no  PBH domination} (i.e., the first case listed above), the GW spectral energy density\footnote{GWs at second order can be sourced by the density fluctuation due to the inhomogeneities in the PBH distribution, which puts a constraint on $\beta$, requiring sub-dominant contribution from GW energy density~\cite{Papanikolaou:2020qtd,Domenech:2020ssp,Domenech:2021wkk,Domenech:2021ztg}.  In our case, we are mainly interested in the scenario with no PBH domination, $\beta<\beta_c$, where such induced GWs spectrum is sub-dominant. Even for $\beta>\beta_{c}$, we chose $\beta$ value close to $\beta_c$, where such induced gravitational waves can be neglected. } at the present epoch can be represented in a piecewise function of frequency (momenta $k$) as follows 
\begin{align}\label{eq:gw-inf}
& \Omega_{\rm GW}^{(0)}\simeq \Omega_{\rm GW,rad}^{(0)}
\begin{cases}
1 & k<\krh
\\[8pt]    
\frac{\zeta}{\pi}\,\left(\frac{k}{\krh}\right)^\frac{6\,w_\phi-2}{1+3\,w_\phi} & \krh <k<k_{\rm max},
\end{cases}
\end{align}
where $\Omega_{\rm GW,rad}^{(0)}h^2
= \frac{\Omega_{\rm R} h^2 H_{\rm end}^2}{12 \pi^2\, M_P^2}$ and 
\begin{align}
\zeta=(1+3\,w_\phi)^\frac{4}{1+3\,w_\phi}\,\Gamma^2\left(\frac{5+3\,w_\phi}{2+6\,w_\phi}\right)\,.   
\end{align}
$\Omega_{\rm R} h^2= 4.16\times10^{-5}$ is the present-day radiation abundance considering both photons and neutrinos. 
Using the entropy conservation between the 
end of reheating to the present day, 
we have the mode that re-enters the Hubble radius at the end of reheating 
\begin{align}
\krh\equiv \arh\,\Hrh = \left(\frac{43}{11\,g_{\rm RH}}\right)^{\frac{1}{3}}\sqrt{\frac{\alpha_T}{3}}\frac{T_{\rm 0}}{M_P}\,T_{\rm RH}\,,
\end{align}
where $T_{\rm 0}$ is the present CMB temperature 2.725. Thus, $k_{\rm RH}$ is simply a function of $T_{\rm RH}$, which can be written as
\beq
k_{\rm RH}\sim 1.6\, \text{Hz} \left(\frac{\trh}{10^7 \,\text{ GeV}}\right)\left(\frac{g_{\rm RH}}{106.75}\right)^{\frac{1}{6}}\,.
\eeq
Since here we are interested in the scenario of PBH reheating without PBH domination, $T_{\rm RH}$ is a function of PBH parameters such as formation mass $M_{\rm in}$ and mass fraction $\beta$ and takes, as we saw, the following form [cf. Eq.~\eqref{Eq:pbhreheattempbis}]

\beq \label{Eq:Pbhreheat}
T_{\rm RH}\sim \mu\,\beta^{\frac{3\,(1+w_\phi)}{4\,(3\,w_\phi-1)}} \left(\frac{M_P}{\Min}\right)^{\frac{3\,(1-w_\phi)}{2\,(1-3\,w_\phi)}}\,M_P\,,
\eeq
where
\begin{align}
& \mu=\left(\frac{48\,\pi^2}{\alpha_T}\right)^{\frac{1}{4}}\left(\frac{\epsilon}{2\,(1+w_\phi)\,\pi\,\gamma^{3w_\phi}}\right)^\frac{1}{2\,(1-3\,w_\phi)}\,.    
\end{align}
On the other hand, the mode re-entering right at the end of inflation is designated as $\kmax$, where
\beq
\kmax=\amax\,H_{\rm end}\simeq k_\star\,e^{N_\star}\,,
\eeq
where $\star$ quantities are measured at the CMB pivot scale $k_\star\simeq 0.05\,\text{Mpc}^{-1}$ and $N_{\rm *}$ represents the inflationary e-folding number calculated from the end of the inflation to the horizon exit of the CMB pivot scale. Under the assumption that the comoving entropy density is conserved from the end of the reheating to the present day, the expression for $N_{\rm *}$ takes the following form
\beq
N_{\rm *}=\ln \left[2.5\times 10^{39}\left(\frac{H_{\rm end}}{10^{13}\,\rm GeV}\right)\frac{\rm GeV}{T_{\rm RH}}\right]-N_{\rm RH}\,.
\eeq
In the first case, where  PBH formed and evaporates in an inflaton-dominated background and is responsible for reheating, one can estimate 
\beq
N_{\rm RH}\simeq \frac{1}{3\,(1+w_\phi)}\ln \left[\frac{H_{\rm end}}{10^{13}\,\rm GeV}\left(\frac{1.5\times 10^{12}\,\rm GeV}{T_{\rm RH}}\right)^4 \right]\,.
\eeq

On the other hand, if there is an intermediate epoch of PBH domination before the reheating ends 
(i.e., the second case), the GW spectrum shows a 
red-tilted behavior ($\propto k^{-2}$) for all GW momenta modes that re-enter the horizon during the 
period $\kbh<k<\krh$. The final GW spectral energy density at the present epoch then takes the form
\begin{align}\label{eq:gw-pbh}
& \Omega_{\rm GW}^{(0)}\simeq \Omega_{\rm GW,rad}^{(0)}
\begin{cases}
1 & k<\krh 
\\[8pt]
c_1\,\left(\frac{k}{k_{\rm RH}}\right)^{-2}& k_{\rm BH}<k<\krh
\\[8pt]
c_2\,\left(\frac{k}{k_{\rm BH}}\right)^\frac{6\,w_\phi-2}{1+3\,w_\phi} & k_{\rm BH}<k<k_{\rm max}
\\[8pt]
\end{cases}
\end{align}
where $c_1=\left[\Gamma\left(\frac{5}{2}\right)\right]^2/\pi$, $c_2=(c\,\zeta/\pi))\,\left(k_{RH}/k_{\rm BH}\right)^2$, and $k_{\rm BH}=k_{\rm max}\left(a_{\rm BH}/\aend\right)^{-(1+3\,w_\phi)/2}$.
\begin{figure*}[t!]
\includegraphics[scale=0.42]{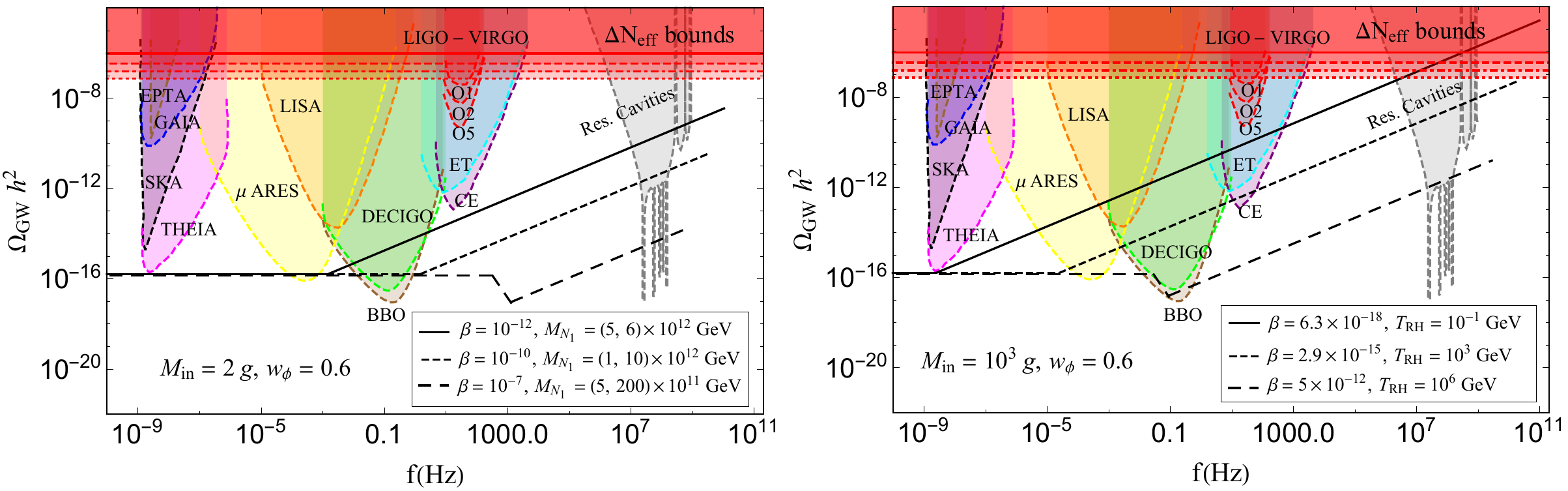}
\caption{{\it Left:} Spectrum of primordial GW as a function of the frequency $f$ shown via the black curves, for different choices of $\beta$ and $M_{N_1}$ that satisfy the observed baryon asymmetry. We fix $\Min=2$ g and $w_\phi=0.6$. {\it Right:} Same as left, but for $\Min=10^3$ g and $w_\phi=0.6$. In both plots, we also show projections from present and future GW experiments, together with the existing and projected $\DNeff$ bounds (labeled as ``$\DNeff$ bounds"). 
}
\label{fig:spectrum}
\end{figure*}

In Fig.~\ref{fig:spectrum}, we show the spectrum of primordial GW as a function of the frequency, along with the current and future sensitivities of various GW experiments, like, LIGO~\cite{LIGOScientific:2016aoc,LIGOScientific:2016sjg,LIGOScientific:2017bnn,LIGOScientific:2017vox,LIGOScientific:2017ycc,LIGOScientific:2017vwq}, LISA~\cite{2017arXiv170200786A,Baker:2019nia}, CE~\cite{LIGOScientific:2016wof,Reitze:2019iox}, ET~\cite{Punturo:2010zz,Hild:2010id}, BBO~\cite{Crowder:2005nr,Corbin:2005ny}, DECIGO~\cite{Seto:2001qf,Kudoh:2005as,Nakayama:2009ce,Yagi:2011wg,Kawamura:2020pcg}, $\mu$-ARES~\cite{Sesana:2019vho} and THEIA~\cite{10.3389/fspas.2018.00011}\footnote{Here we have used the sensitivity curves derived in Ref.~\cite{Schmitz:2020syl}.},
that search for signals in the low frequency (kHz) regions. We also project sensitivity from proposed high frequency GW experiments, e.g., resonant cavities~\cite{Ringwald:2020ist,Ringwald:2022xif} that typically look for GW signals in GHz-MHz frequency regime.

In the left panel of Fig.~\ref{fig:spectrum} we show GW spectrum corresponding to different choices of $\beta$, for a fixed $M_{N_1}$ with a given $\Min=2$ g and $w_\phi=0.6$. Note that, for these choices of the RHN masses one can satisfy the observed baryon asymmetry by exploiting the Casas-Ibarra parametrization. The noteworthy feature here is the scale-invariant spectrum, followed by a blue-tilted branch in case of $\beta<\beta_c$. This, as explained before, is because of the presence of stiff equation of state during reheating, when PBH domination is absent. For $\beta>\beta_c$, we see the effect of intermediate PBH-domination (black-dashed curve) that gives rise to red-tilted spectrum. In the right panel, we see similar behaviour of the GW spectrum, but now with different choices of $\trh$, for $\Min=10^3$ g and $w_\phi=0.6$. In both panel the horizontal lines, along with the shaded region marked as ``$\DNeff$ bound" collectively shows present and future bounds from $\DNeff$ from different experiments which we are going to explain in the very next section. 
\subsection*{Constraints from $\Delta N_{\rm eff}$}
Any extra radiation component, in addition to those of the SM, can be expressed in terms of the $\DNeff$. This can be done by computing the total radiation energy density in the late Universe as
\begin{align}
&\rho_\text{rad} = \rho_\gamma + \rho_\nu + \rho_\text{GW} 
\nonumber\\&
= \left[1 + \frac78 \left(\frac{T_\nu}{T_\gamma}\right)^4 N_\text{eff}\right] \rho_\gamma\,,
\end{align}
where $\rho_\gamma$, $\rho_\nu$, and $\rho_\text{GW}$ correspond to the photon, SM neutrino, and GW energy densities, respectively, with $T_\nu/T_\gamma = (4/11)^{1/3}$. Within the SM, taking the non-instantaneous neutrino decoupling into account, one finds $N_\text{eff}^\text{SM} = 3.044$~\cite{Dodelson:1992km, Hannestad:1995rs, Dolgov:1997mb, Mangano:2005cc, deSalas:2016ztq, EscuderoAbenza:2020cmq, Akita:2020szl, Froustey:2020mcq, Bennett:2020zkv}, while the presence of GW results in a modification 
\begin{align}
& \DNeff = N_\text{eff}-N_\text{eff}^\text{SM} = \frac{8}{7}\,\left(\frac{11}{4}\right)^\frac{4}{3}\,\left(\frac{\rho_\text{GW}(T)}{\rho_\gamma(T)}\right)\,.
\end{align}
The above relation can be utilized to put a constraint on the GW energy density red-shifted to today via~\cite{Maggiore:1999vm,Boyle:2007zx,Caprini:2018mtu}
\begin{align}\label{Eq:BBNgw}
  \int_{k_{_\text{BBN}}}^{k_{\rm max}}\frac{dk}{k}\Omega_{_{\rm GW}}^{(0)}\,h^2(k)\leq \frac{7}{8}\left(\frac{4}{11}\right)^{4/3}\Omega_{\rm \gamma}h^2\,\DNeff\,,
\end{align}
where $\Omega_{\rm \gamma}h^2\simeq 2.47\times10^{-5}$ is the relic density of the photon today. 
\begin{table}[t!]
    \begin{center}
        \begin{tabular}{|c||c|}
            \hline
            $\DNeff$ & Experiments  \\ 
            \hline\hline
            $0.17$ & Planck legacy data (combining BAO)~\cite{Planck:2018vyg}\\
            $0.14$ & BBN+CMB combined~\cite{Yeh:2022heq}\\
            $0.06$ & CMB-S4~\cite{Abazajian:2019eic} \\
            $0.027$ & CMB-HD~\cite{CMB-HD:2022bsz}\\
            $0.013$ &  COrE~\cite{COrE:2011bfs}, Euclid~\cite{EUCLID:2011zbd} \\
            $0.06$ & PICO~\cite{NASAPICO:2019thw} \\
            \hline
        \end{tabular}
    \end{center}
    \caption {Present and future constraints on $\DNeff$ from different experiments.}
    \label{tab:DNeff}
\end{table}

The $\DNeff$ constraints on the GW spectral energy density become relevant when the PBHs evaporate in an inflaton-dominated background. Therefore, for a background equation of state $w_\phi>1/3$, the blue tilted nature of the spectrum becomes apparent with maximum momenta $k_{\rm max}$ for the mode that re-enters the horizon right at the end of the inflation. In this case, Eq.~\eqref{Eq:BBNgw} takes the form
\begin{align}\label{Eq:BBNapprox}
  \int_{k_{_\text{BBN}}}^{k_{\rm max}}\frac{dk}{k}\Omega_{\rm GW}^{(0)}\,h^2(k)\simeq \Omega_{\rm GW,rad}^{(0)}\,h^2 \mu\,\left(\frac{k_{\rm max}}{k_{\rm RH}}\right)^\frac{6\,w_\phi-2}{1+3\,w_\phi}
\end{align}
where $\mu=\frac{\zeta\,(1+3\,w_\phi)}{2\pi\,(3\,w_\phi-1)}$. Assuming a $w_\phi$ dominated phase between inflation and radiation domination (no PBH domination), the ratio between $k_{\rm max}$ and $k_{\rm RH}$ can be expressed as 
\bea
\frac{k_{\rm max}}{k_{\rm RH}}=\left(\frac{\rho_{\rm end}}{\alpha_T}\right)^{\frac{1+3\,w_\phi}{6\,(1+w_\phi)}}\,T_{\rm RH}^{-\frac{2}{3}\frac{(1+3\,w_\phi)}{(1+w_\phi)}}.
\eea
Upon substitution of the above equation into Eq.~\eqref{Eq:BBNapprox} and utilizing Eq.~\eqref{Eq:BBNgw}, one can find the restriction on the reheating temperature
\bea \label{Eq: PBH }
T_{\rm RH}\geq \left(\frac{\Omega_{\rm GW,rad}^{(0)}h^2 \mu}{5.61\times 10^{-6}\,\DNeff}\right)^{\frac{3\,(1+w_\phi)}{4\,(3\,w_\phi-1)}}\times \left(\frac{\rho_{\rm end}}{\alpha_T}\right)^{\frac{1}{4}}\,,
\eea
where $\rho_{\rm end}$ is the energy density of the inflaton at the end of the inflation, $\rho_{\rm end}=3\,M_P^2\,H_{\rm end}^2$. Since we are interested in the PBH reheating scenario (no PBH domination), the above restriction on the reheating temperature, in turn, puts bounds on the PBH parameter (cf. Eq.~\eqref{Eq:Pbhreheat} via 
\begin{align}\label{eq:beta-bound}
&\beta\geq \left(\frac{\Omega_{\rm GW,rad}^{(0)}h^2 \mu}{5.61\times 10^{-6}\,\DNeff}\right)\left(\frac{M_P}{M_{\rm in}}\right)^{\frac{2\,(1-w_\phi)}{1+w_\phi}}\nonumber
\\& \times \left(\frac{1}{\mu^4\,\alpha_{\rm T}}\frac{\rho_{\rm end}}{M_P^4}\right)^{\frac{3\,w_\phi-1}{3\,(1+w_\phi)}}\,.
\end{align}
Note that the above restriction is only important when $T_{\rm RH}>T_{\rm BBN}=4\,\rm MeV$ \cite{Kawasaki:1999na,Kawasaki:2000en,Hasegawa:2019jsa}, otherwise, BBN provides a stronger bound than $\DNeff$. In Tab.~\ref{tab:DNeff}, we tabulate present and future bounds on $\DNeff$ from different experiments as mentioned. These bounds are projected in Fig.~\ref{fig:spectrum}, from where we see that lower reheating temperatures are typically in conflict with these bounds, as seen from the right panel of Fig.~\ref{fig:spectrum}.   
\begin{figure}
\includegraphics[scale=0.38]{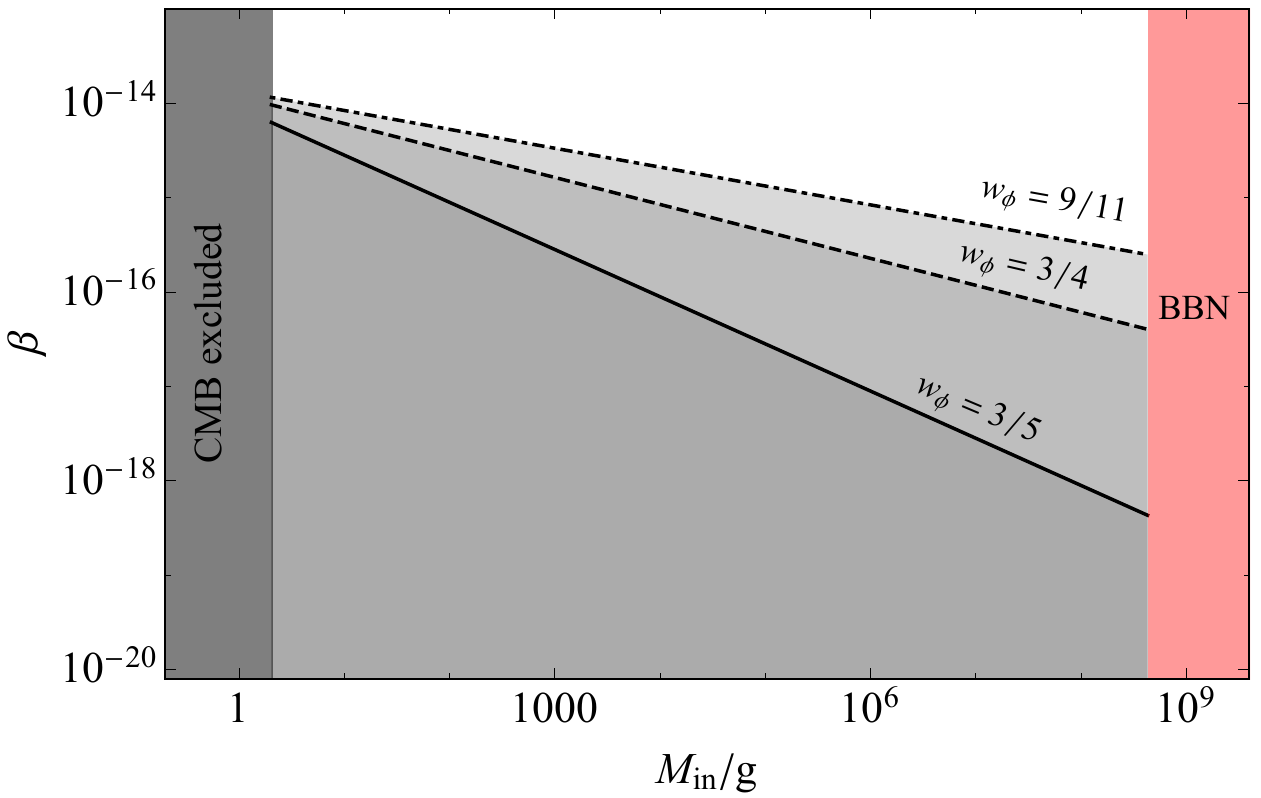}
\caption{Constraint from $\DNeff$ in $\beta-\Min$ plane for different choices of the background equation of states $w_\phi$. All shaded regions are excluded (see text for details).}
\label{fig:neff}
\end{figure}
The $\DNeff$ bound on the translated into a bound on PBH mass and $\beta$-value following Eq.~\eqref{eq:beta-bound}.
This is shown in Fig.~\ref{fig:neff}, where we use the present bound on $\DNeff$ from Planck~\cite{Planck:2018vyg}. We see the available parameter space is more tightly constrained for heavier PBH in the case of a stiffer fluid (i.e., a larger equation of state). As one can see from Eq.~\eqref{eq:beta-bound}, the bound becomes independent of PBH mass as $w_\phi\to 1$, i.e., pure kination. Note that the $\DNeff$ bound for the contribution from the PGWs is only important when $w_\phi>0.60$ ($n>8$)\cite{Chakraborty:2023ocr}. Thus, our presented results are safe from such restrictions.
\section{Conclusions}
\label{sec:concl}
The observed baryon asymmetry of the Universe can be produced via  
leptogenesis, which requires extending the Standard Model (SM) particle 
content with the addition of right-handed neutrinos (RHN), singlet under 
the SM gauge symmetry. 
Gravitation should produce a minimal
unavoidable amount of RHN fields. Once gravitationally produced, these RHNs 
can then undergo CP-violating decay to produce the lepton and 
subsequently, the baryon asymmetry. 

Pure gravitational production can 
take place in two ways: (i) from evaporation of primordial black holes 
(PBH) and (ii) 
scattering of the inflaton (or bath particles), mediated by a
massless graviton field. The latter production occurs during reheating when the inflaton field $\phi$ 
oscillates around the minima of a monomial potential $V(\phi)\propto\phi^n$, transferring its energy to the thermal bath. In the presence of PBHs, however, 
the reheating dynamics is controlled not only by the steepness of the 
potential $n$ and the nature of the inflaton-SM coupling, but also by the 
PBH mass and the fractional abundance of PBHs. Moreover, for $n>4$ (or 
equivalently, a general equation of state $w_\phi>1/3$), the primordial 
gravitational waves produced from the tensor perturbations during 
inflation, are hugely blue-tilted. Such a boosted GW energy density on 
one hand falls within the sensitivity range of GW detectors, while on the 
other hand, may also be in tension with excessive production of energy 
density around BBN.

In the present work, we first compute
the amount of baryonic asymmetry 
$Y_B^0$ at the present epoch, generated through the gravitational production of RHN. Depending 
on the relative amount of energy
$\beta$, the PBHs
can lead the reheating process
and produce an amount of RHN sufficient to satisfy the constraint
on $Y_B^0$. However, this is possible only for very light PBH $\lesssim 10$ g, as one can see in Fig.~(\ref{fig:fig1}). If PBHs do not dominate the energy budget of the Universe at the time of reheating, the situation 
worsens due to an excessive entropy dilution, as it is clear from Fig.~(\ref{fig:lep-beta}), plotted for smaller values of $\beta$. For a complete picture, one has also included
the other gravitational source of RHNs, i.e., the inflation scattering through graviton exchange. In this scenario, a new region of the parameter space opens up with larger PBH masses, as one can see from Fig.~(\ref{fig:leptograv}). In any case, 
the allowed mass range for the RHN remains
$5\times 10^{11}~\text{GeV} \lesssim M_{N_1} \lesssim 5\times 10^{14}~\rm{GeV}$. Too-light RHNs do not generate a sufficient amount of asymmetry, while too-heavy RHNs are not sufficiently produced by inflaton scattering or from PBH  evanescence. 

It is then possible to find signatures of different reheating, as well as gravitational production scenarios (inflaton or PBH sourced) through GW observations. 
We exploit the blue-tilted nature of primordial GW in 
probing the scale of non-thermal gravitational leptogenesis during 
reheating. The reheating via inflaton is controlled by the Yukawa coupling between 
the inflaton and a pair of SM-like fermions, {\it viz.,} 
$y_\phi\,\phi\,f\,\bar f$. PBHs, however, are assumed to be formed during 
the epoch of reheating and are 
parameterized by their formation mass $\Min$ and initial abundance $\beta$. Depending 
on the values of $\{y_\phi,\,\Min,\,\beta,\,n\}$, PBHs 
can potentially impact the reheating process and populate the thermal bath. In Fig.~(\ref{fig:spectrum}), we delineate the parameter 
space that agrees with the observed 
baryon asymmetry, considering RHN production takes place during 
reheating both from PBH evaporation and from the scattering of inflaton 
condensate (mediated by graviton).

For a stiff equation of state for the background $\phi$ (large $n$), we observe that the spectrum 
of primordial GW lies well within reach of future GW detectors [cf.Fig.~(\ref{fig:spectrum})], both in the low frequency (kHz) and in the high frequency (GHz) regime, 
satisfying bounds from $\DNeff$, as one can see from Fig.~(\ref{fig:spectrum}) and (\ref{fig:neff}). Interestingly, the {\it red-tilted} GW 
spectrum that exists because of the 
intermediate PBH domination (for $\beta>\beta_c$), also turns out to 
be within the reach of futuristic GW detectors. The present scenario 
therefore provides a window to test modified cosmological background 
prior to BBN, induced by inflaton and PBH dynamics, together with purely 
gravitational leptogenesis through 
primordial GW spectrum. 
\acknowledgements
This project has received support from the European Union's Horizon 2020 research and innovation program under the Marie Sklodowska-Curie grant agreement No 860881-HIDDeN and the CNRS-IRP project UCMN. BB would like to acknowledge the ``Planck 2023" conference at the University of Warsaw, during which discussions regarding this work was initiated. SJD is supported by IBS under the project code IBS-R018-D1. SJD would also like to thank the support and hospitality of the Tata Institute of Fundamental Research (TIFR), Mumbai, where a part of this project was carried out. MRH wishes to acknowledge support from the Science and Engineering Research Board (SERB), Government of India (GoI), for the SERB National Post-Doctoral fellowship, File Number: PDF/2022/002988. 
\appendix
\section{The Boltzmann equations for PBH-inflaton-radiation system}
\label{sec:BEQ}
\begin{figure*}[htp]
\centering
\subfigure[]{\includegraphics[scale=0.38]{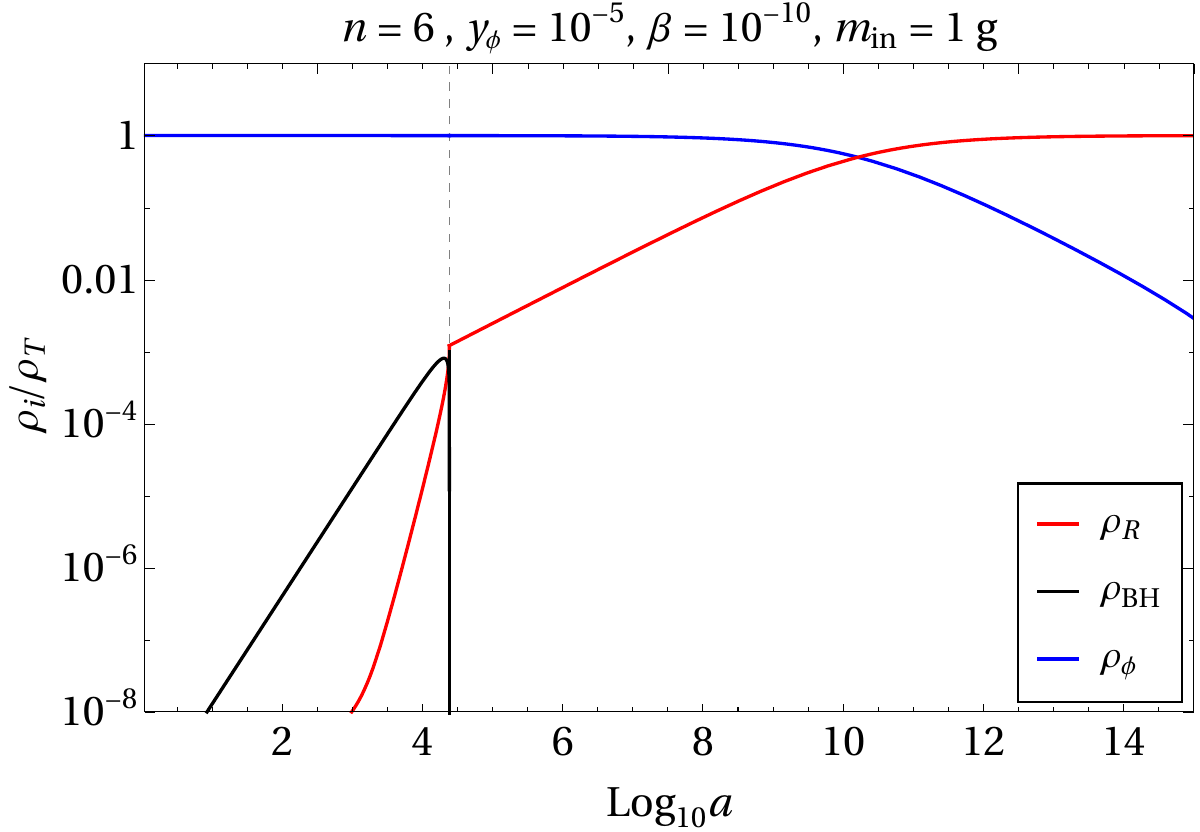}}
\quad
\subfigure[]{\includegraphics[scale=0.38]{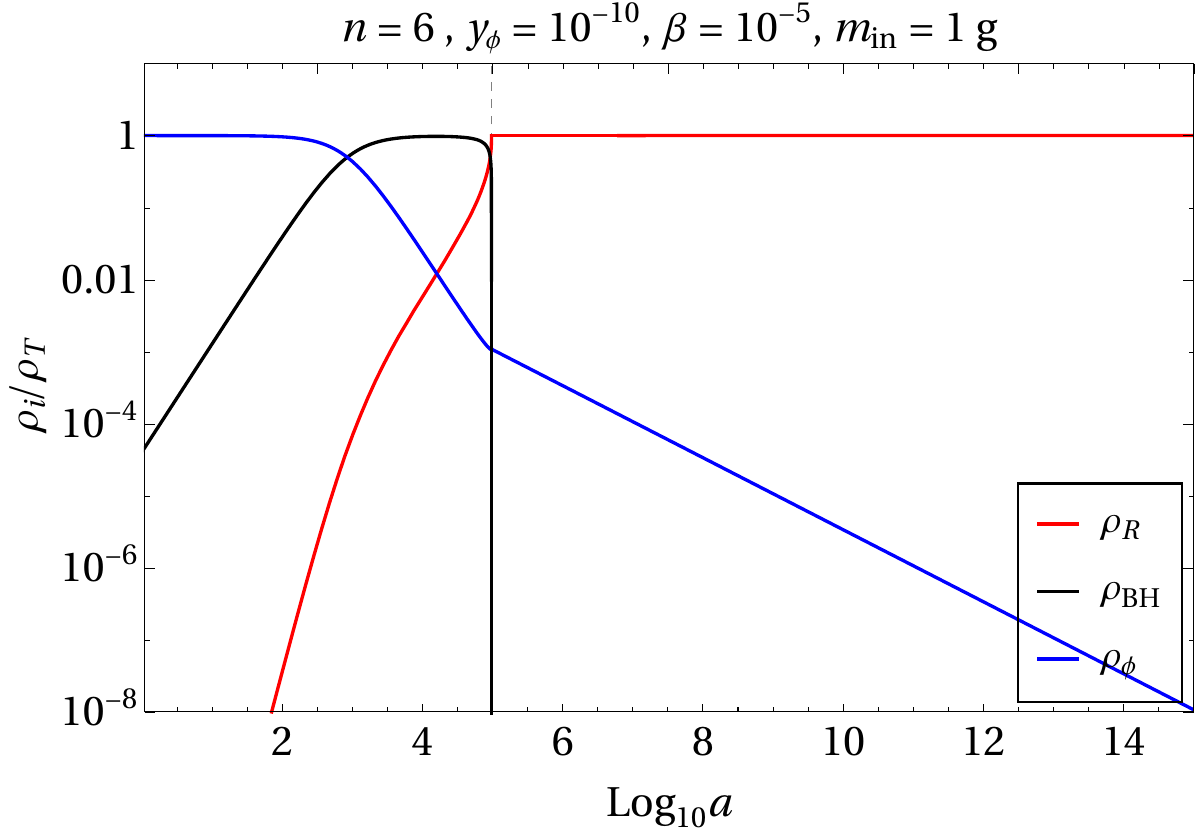}}
\quad
\subfigure[]{\includegraphics[scale=0.38]{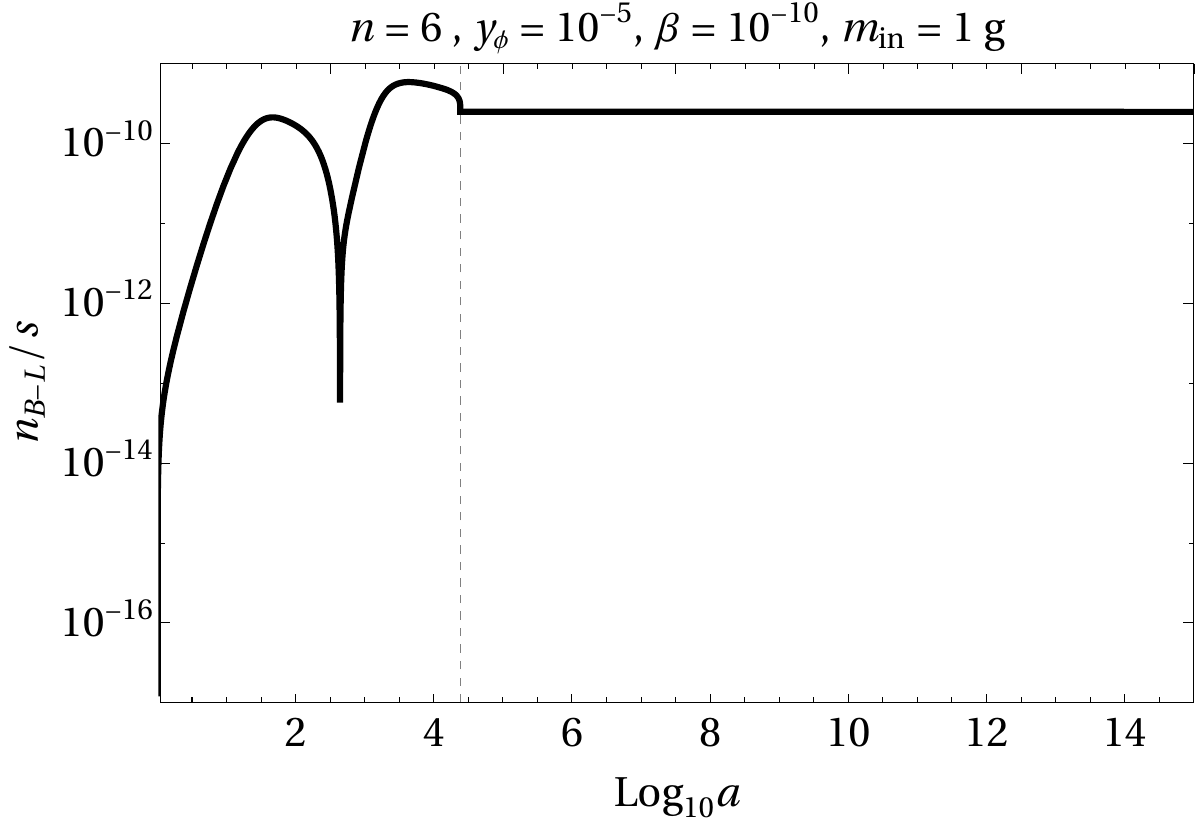}}
\quad
\subfigure[]{\includegraphics[scale=0.38]{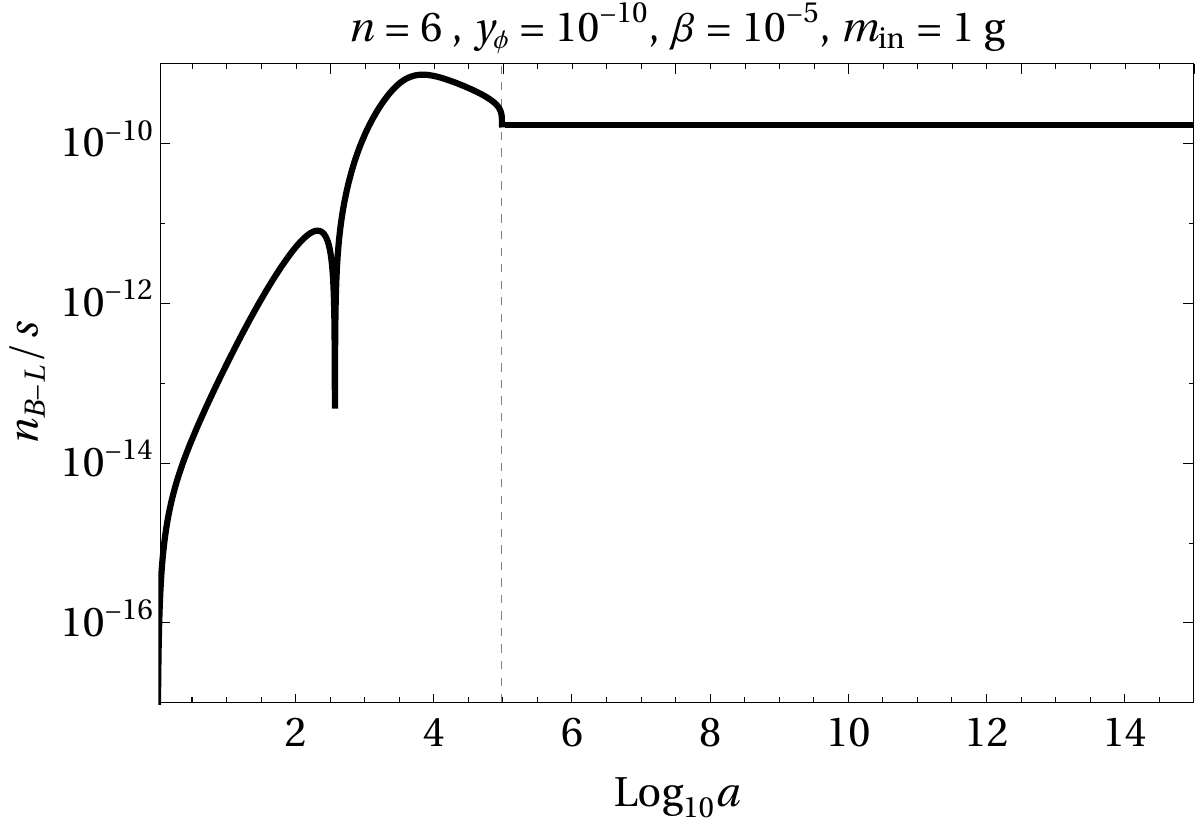}}
\caption{{\it Top:} Evolution of radiation (red), inflaton (blue) and PBH (black) energy densities with the scale factor, obtained by numerically solving Eq.~\eqref{eq:beq}. {\it Bottom:} Evolution of $n_{B-L}/s$ as a function of the scale factor. The final asymmetry satisfies the observed value $Y_B^0\simeq 8.7\times 10^{-11}$. All relevant parameters are mentioned in the plot legend.} 
\label{fig:rhoplt1}
\end{figure*}

In order to track the evolution of radiation ($\rho_R$), PBH ($\rbh$) and inflaton ($\rho_\phi$) energy densities, together with number density of right handed neutrinos ($n_N$), the asymmetry $B-L$ and the Hubble parameter $H$, we solve the following set of Boltzmann equations numerically
\begin{align}\label{eq:beq}
&\frac{d\rho_\phi}{da}+3\,(1+w_\phi)\frac{\rho_\phi}{a}=-\frac{\Gamma_\phi}{H}\,\left(1+w_\phi\right)\,\frac{\rho_\phi}{a}\,,
\nonumber
\\&
\frac{d{\rho}_R}{da}+4\,\frac{\rho_R}{a}=-\frac{\rho_\text{BH}}{M_\text{BH}}\,\frac{dM_\text{BH}}{da}\, 
+\frac{\Gamma_\phi\,\rho_\phi\,(1+w_\phi)}{a\,H},
\nonumber\\&
\frac{d{\rho}_\text{BH}}{da}+3\,\frac{\rho_{\rm BH}}{a}=\frac{\rho_{\rm BH}}{M_\text{BH}}\,\frac{dM_\text{BH}}{da}\,,
\nonumber
\\&
\frac{dn^{\rm BH}_{N_1}}{da}+3\frac{n^{\rm BH}_{N_1}}{a}=-n^{\rm BH}_{N_1}\,\Gamma_{N_1}^{\rm BH}+\Gamma_{\text{BH}\to N_1}\,\frac{{\rho}_\text{BH}}{M_\text{BH}}\,\frac{1}{a\,H}\,,
\nonumber
\\&
\frac{d{n}_{B-L}}{da}+3\,\frac{n_{B-L}}{a}= \frac{\kappa_{\Delta L}}{a\,H}\Big[\left(n_{N_1}^T-n_{N_1}^{\rm eq}\right)\,\Gamma_{N_1}^T+n_{N_1}^{\rm BH}\,\Gamma_{N_1}^{\rm BH}\Big]\,,
\nonumber
\\&
\frac{dM_{\rm BH}}{da} = - \epsilon \frac{M^4_P}{M_{\rm BH}^2}\,\frac{1}{a\,H}\,,
\\&
H^2=\frac{\rho_\phi+\rho_R+\rho_{\rm BH}}{3\,M_P^2}\,.
\nonumber
\end{align}
Here the `T' and `BH' stands for thermal and PBH (non-thermal) contributions respectively. We define $\Gamma_N^{\rm BH}$ as the
decay width corrected by an average time dilation factor
\begin{align}
&  \Gamma_N^{\rm BH} = \Big\langle\frac{M_N}{E_N}\Big\rangle_{\rm BH}\,\Gamma_N\approx \frac{K_1(M_N/\Tbh)}{K_2(M_N/\Tbh)}\,\Gamma_N\,, 
\end{align}
where $K_{1,2}[...]$ are the modified Bessel functions of second kind and the thermal average is obtained assuming that the Hawking spectrum has a Maxwell-Boltzmann form, while
\begin{align}
& \Gamma_N = \frac{M_N}{8\,\pi}\,y_N^\dagger\,y_N    
\end{align}
is the total decay RHN decay width, with $y_N$ being parametrized following Eq.~\eqref{eq:CI}. Here $\Gamma_{{\rm BH}\to N_1}$ is the non-thermal production term for RHNs (originating from PBH evaporation)
and can be written as
\begin{align}
& \Gamma_{{\rm BH}\to N_1} = \int \frac{d^2\,\mathcal{N}}{dp\,dt}\,dp
\nonumber\\&
\approx\frac{27\,\Tbh}{32\,\pi^2}\,\Big[-z_{\rm BH}\,\text{Li}_2\left(-e^{-z_{\rm BH}}\right)-\text{Li}_3\left(-e^{-z_{\rm BH}}\right)\Big]\,,
\end{align}
where $\text{Li}_s[...]$ are polylogarithm functions of order $s$; assuming the greybody factor equal to the geometric optics limit, such analytical expression is obtained. In Fig.~\ref{fig:rhoplt1} we show the evolution of energy densities and the $B-L$ asymmetry as a function of the scale factor for two benchmark values of $y_\phi$ and $\beta$ such that in one case (top left panel) inflaton dominates the reheating process, while in the other (top right panel) it is dominated by the PBH.

\section{Casas-Ibarra Parametrization}
\label{sec:CI}
As the neutral component of the SM Higgs doublet acquires a VEV leading to the spontaneous breaking of the SM gauge symmetry, neutrinos in the SM obtain a Dirac mass that can be written as
\bea
m_D= \frac{y_{N}}{\sqrt{2}}v.
\eea
The Dirac mass $m_D$ together with the RHN bare mass $M_N$, can explain the nonzero light neutrino masses with the help of Type-I seesaw~\cite{GellMann:1980vs, Mohapatra:1979ia,MINKOWSKI1977421}. Here, the light-neutrino masses can be expressed as,
\bea
m_{\nu}\simeq m_{D}^T~M^{-1}~m_{D}.
\eea
The mass eigenvalues and mixing are then obtained by diagonalizing the light-neutrino mass matrix as
\bea
m_{\nu}=\mathcal{U}^*\,m_{\nu}^d\, \mathcal{U}^{\dagger}\,,
\eea
with $m_{\nu}^d=\text{diag}\{m_{\nu_1},\,m_{\nu_2},\,m_{\nu_3}\}$, consisting of the mass eigenvalues and $\mathcal{U}$ being the Pontecorvo-Maki-Nakagawa-Sakata matrix~\cite{Zyla:2020zbs}\footnote{The charged lepton mass matrix is considered to be diagonal.}. In order to obtain a complex structure of the Yukawa coupling which is essential from the perspective of leptogenesis, we use the well-known Casas-Ibarra (CI) parametrisation~\cite{Casas:2001sr}. Using this one can write the Yukawa coupling $y_N$ as,
\bea
y_N = \frac{\sqrt{2}}{v}\sqrt{M}~\mathbb{R}~\sqrt{m_{\nu}^d}~\mathcal{U}^{\dagger}\,,
\label{eq:CI}
\eea
\noindent where $\mathbb{R}$ is a complex orthogonal matrix $\mathbb{R}^T \mathbb{R} = I$, which we choose as
\begin{align}
\mathbb{R} =
\begin{pmatrix}
0 & \cos{z} & \sin{z}\\
0 & -\sin{z} & \cos{z}
\end{pmatrix}\,,
\label{eq:rot-mat}
\end{align} 
where $z=a+ib$ is a complex angle. The digonal light neutrino mass matrix $m_{\nu}^d$ is calculable using the best fit values obtained from the latest neutrino oscillation data~\cite{Zyla:2020zbs}. The elements of Yukawa coupling matrix $y_N$ for a specific value of $z$, can be obtained for different choices of the heavy neutrino masses.
\section{Expression for the CP asymmetry}
\label{sec:cp}
The CP asymmetry generated from $N_1$ decay is given by~\cite{Davidson:2008bu}
\begin{align}\label{eq:cp1}
& \kappa_{\Delta L} \equiv \frac{\Gamma_{N_1 \to \ell_i\, H } -\Gamma_{N_1 \to \bar\ell_i\, \bar H}}{\Gamma_{N_1 \to \ell_i\, H} + \Gamma_{N_1 \to \bar\ell_i\, \bar H}} \simeq \frac{1}{8\, \pi}\, \frac{1}{(y_N^\dagger\, y_N)_{11}}
\nonumber\\&
\sum_{j=2, 3} \text{Im}\left(y_N^\dagger\, y_N\right)^2_{1j}
\times \mathcal{F}\left(\frac{M_j^2}{M_1^2}\right),
\end{align}
where
\begin{equation}
    \mathcal{F}(x) \equiv \sqrt{x}\,\left[\frac{1}{1-x}+1-(1+x)\,\log\left(\frac{1+x}{x}\right)\right].
\end{equation}
For $x\gg 1\,,\mathcal{F}\simeq -3/\left(2\,\sqrt{x}\right)$, and Eq.~\eqref{eq:cp1} becomes
\begin{align}
&\kappa_{\Delta L} \simeq -\frac{3}{16\, \pi}\, \frac{1}{(y_N^\dagger\, y_N)_{11}} \Big[\text{Im}\left(y_N^\dagger\, y_N\right)^2_{12} \frac{m_{N_1}}{m_{N_2}} 
\nonumber\\&
+ \text{Im}\left(y_N^\dagger\, y_N\right)^2_{13} \frac{m_{N_1}}{m_{N_3}}\Big].
\end{align}
If we assume $\text{Im}\left(y_N^\dagger\, y_N\right)^2_{13} \gg \text{Im}\left(y_N^\dagger\, y_N\right)^2_{12}$ and $m_{N_1}\ll m_{N_{2,3}}$, then 
\begin{equation}
    \kappa_{\Delta L} \simeq -\frac{3\, \delta_\text{eff}}{16\, \pi}\,\frac{|(y_N)_{13}|^2 \, m_{N_1}}{m_{N_3}}\,,   
\end{equation}
while the effective CP violating phase is given by
\begin{equation}
    \delta_\text{eff} = \frac{1}{(y_N)_{13}^2}\, \frac{\text{Im}(y_N^\dagger\,y_N)^2_{13}}{(y_N^\dagger\,y_N)_{11}}\,.    
\end{equation}
In order to simultaneously generate the active neutrino mass, one has to impose the seesaw relation
\begin{equation}
    m_{\nu_3} = \frac{|(y_N)_{13}|^2\, v^2}{m_{N_3}}\,,    
\end{equation}
that leads to
\begin{equation}
    \kappa_{\Delta L} \simeq -\frac{3\, \delta_\text{eff}}{16\, \pi}\, \frac{m_{N_1}\, m_{\nu_3}}{v^2}\,.
\end{equation}
Instead, if $\text{Im}\left(y_N^\dagger\, y_N\right)^2_{13} \ll \text{Im}\left(y_N^\dagger\, y_N\right)^2_{12}$, the CP asymmetry
parameter becomes
\begin{equation}
    \kappa_{\Delta L} \simeq -\frac{3\, \delta_\text{eff}}{16\, \pi}\, \frac{m_{N_1}\, m_{\nu_2}}{v^2}\,.    
\end{equation}
In general, one can then write
\begin{equation}
    \kappa_{\Delta L} \simeq -\frac{3\, \delta_\text{eff}}{16\, \pi}\, \frac{m_{N_1}\, m_{\nu_i}}{v^2}\,,    
\end{equation}
where $i=2\,,3$ for normal hierarchy. In a similar fashion, the CP-asymmetry parameter can be obtained for the inverted hierarchy with $i=1\,,2$. 
\section{Details of the reheating dynamics}
\label{sec:inf}
Considering the inflaton-SM interaction of the form $y_\phi\phi\bar{f}f$, where $f$ are the SM-like fermions, we have the radiation energy density as
\bea
\rho_R^{\rm D}(a) &=& \frac{y_\phi^2}{8 \pi}\, \lambda^{\frac{1-w_\phi}{2\,(1+\,w_\phi)}}  \tilde{\alpha}_n\,M_P^4
\left(\frac{\rhoe}{M_P^4}\right)^{\frac{3}{2}-\frac{1}{1+w_\phi}} \left(\frac{a}{\ae}\right)^{-4}
\nonumber
\label{Eq:rhorwobh}
\\
&&
\times
\left[\left(\frac{a}{\ae}\right)^{\frac{5-9\,w_\phi}{2}}-1\right] \,,
\eea
and
\beq
\tilde{\alpha}_n=\frac{\sqrt{3n^3(n-1)}}{7-n}M_P^4\,,
\eeq
where $\ae$ is the scale factor associated with the end of inflation and $\lambda$ can be expressed in terms of the amplitude of the CMB power spectrum $A_s$ as
\begin{align}\label{eq:lambda}
&\lambda\simeq\frac{18\,\pi^2\,A_s}{6^{n/2}\,N_e^2}\,,    
\end{align}
for the $\alpha$-attractor potential~\cite{Kallosh:2013hoa, Kallosh:2013yoa} of the form
\begin{align} \label{eq: attractor}
&    V(\phi ) =\lambda\, M_P^4 \left[\tanh \left(\frac{\phi}{\sqrt{6\, \alpha}\, M_P}\right)\right]^n 
    \nonumber\\&
    \simeq \lambda\, M_P^4 \times
    \begin{cases}
        1\,, &\phi \gg M_P,\\
        \left(\frac{\phi}{\sqrt{6\,\alpha}\,M_P}\right)^n\,,&\phi\ll M_P\,.
    \end{cases}
\end{align}
Here $N_e$ is the number of e-folds measured from the end of inflation to the time when the pivot scale $k_\star\simeq 0.05\,\text{Mpc}^{-1}$ exits the horizon. In our analysis, we consider $\log(10^{10}\,A_s)=0.04$~\cite{Planck:2018jri} and set $N_e = 55$.

One can find the effective mass of the inflaton which is defined as the second derivative of the inflaton potential as 
\bea \label{Eq: inflaton mass}
m^2_\phi(t)=V''(\phi_0(t))=n\,(n-1)\lambda\, M_P^2\left(\frac{\phi_0(t)}{M_P}\right)^{n-2}.~
\eea
Assuming that the oscillation's time scale is small compared to the decay and redshift time scales, $\phi_0(t)$ captures the impact of both decay and redshift. The inflaton energy density $\rho_\phi=\langle({\dot \phi}^2/2) + V(\phi)\rangle \sim V(\phi_0)$ can be approximated by $\langle\dot{\phi}^2\rangle\simeq \langle\phi \, V'(\phi)\rangle$, that is obtained by averaging the single oscillation. With that assumption, the expression for inflaton mass is obtained as
\bea \label{Eq: inflaton mass-1}
m^2_\phi(t)=n\,(n-1)\lambda^{\frac{2}{n}}\, M_P^2\left(\frac{\rho_\phi}{M_P^4}\right)^{\frac{n-2}{n}}\,.
\eea
    
Defining the end of reheating (onset of radiation domination) as $\rho_\phi(\arh)=\rho_{\rm R} (\arh)=\rho_{\rm RH}$ one finds
\begin{align}
& \frac{\arh}{\ae} = 
\begin{cases}
\left[\frac{y_\phi^2}{8\pi}\tilde{\alpha}_n\left(\frac{\lambda\,M_P^4}{\rho_{\rm end}}\right)^{\frac{1-w_\phi}{2\,(1+w_\phi)}}\right]^{\frac{2}{3\,(w_\phi-1)}}\,,~~n<7
\\
\left[-\frac{y_\phi^2}{8\pi}\tilde{\alpha}_n\left(\frac{\lambda\,M_P^4}{\rho_{\rm end}}\right)^{\frac{1-w_\phi}{2\,(1+w_\phi)}}\right]^{\frac{1}{1-3\,w_\phi}}\,,~~n>7\,,
\end{cases}
\end{align}
that leads to
\begin{align}
&  \rho_{\rm RH}^{\rm D}=
\begin{cases} \label{Eq: reheatinflaton}
\left(\frac{y_\phi^2}{8\,\pi}\,\tilde{\alpha}_n\right)^{\frac{2\,(1+w_\phi)}{1-w_\phi}}\,\mathcal{M}\,,~~n<7
\\
\left(\frac{y_\phi^2}{8\,\pi}\,\tilde{\alpha}_n\right)^{\frac{3\,(1+w_\phi)}{3\,w_\phi-1}}\,\mathcal{M}^{\frac{3\,(1-w_\phi)}{2\,(3\,w_\phi-1)}}\,
\rho_{\rm end}^{\frac{5-9\,w_\phi}{2\,(1-3\,w_\phi)}}\,,~~n>7\,,
\end{cases}
\end{align}
where $\mathcal{M}=\lambda\,M_P^4$.
Now, the radiation energy density at the end of PBH-driven reheating reads
\beq \label{Eq:pbhev}
\rho_{\rm RH}=\rho_R(\aev)\left(\frac{\aev}{\arh}\right)^4\simeq \rho_{\rm BH}(\aev)\left(\frac{\aev}{\arh}\right)^4 \,,
\eeq
while the inflaton energy density reads 
\bea \label{Eq:inflatonev}
\rho_\phi(\arh)=\rho_\phi(\ain)\left(\frac{\ain}{\arh}\right)^{3\,(1+w_\phi)}\,.
\eea
Upon substitution of Eq.(\ref{Eq:bhenergyden}) into Eq.(\ref{Eq:pbhev}) and comparing with (\ref{Eq:inflatonev}), one can find
\bea
\left(\frac{\aev}{\arh}\right)^4=\beta^{\frac{4}{3\,w_\phi-1}}\left(\frac{\ain}{\aev}\right)^{\frac{12\,w_\phi}{1-3\,w_\phi}}\,.
\eea
Utilizing the above equation, $\rho_{\rm RH}$ can be written as,
\bea
&&
\rho_{\rm RH}=48\pi^2\,\beta^{\frac{3\,(1+w_\phi)}{3\,w_\phi-1}}\left(\frac{\epsilon}{2\,(1+w_\phi)\,\pi\,\gamma^{3\,w_\phi}}\right)^\frac{2}{1-3\,w_\phi}
\nonumber
\\
&&
\left(\frac{M_P}{\Min}\right)^{\frac{6\,(1-w_\phi)}{1-3\,w_\phi}}\,M_P^4\,.
\label{Eq:pbhreheattempbis}
\eea
Imposing the condition that for $y_\phi=y_\phi^c$ one should have $\rho_{\rm RH} =\rho_{\rm RH}^{\rm D}$, we find
\begin{align}
& y_\phi^c=
\begin{cases}
\sqrt{\frac{8\pi}{\tilde{\alpha}_n}}\times\beta^{\frac{3\,(1-w_\phi)}{4\,(3w_\phi-1)}}\times\mathcal{A}\,,~~n<7  
\\
\sqrt{-\frac{8\pi\,\beta}{\tilde{\alpha}_n}}\times\left(\frac{\rho_{\rm end}}{M_P^4}\right)^{\frac{5-9\,w_\phi}{12\,(1+\,w_\phi)}}\times\mathcal{B}\,,~~n>7\,,
\end{cases}
\label{Eq:yphic}
\end{align}
where $\mathcal{A}=\left(\frac{48\pi^2}{\lambda}\right)^{\frac{1-w_\phi}{4\,(1+w_\phi)}}\,\mathcal{E}^{\frac{1-w_\phi}{2\,(1-3\,w_\phi)\,(1+w_\phi)}}\,r^{\frac{3}{2}\frac{(1-w_\phi)^2}{(1-3w_\phi)\,(1+w_\phi)}}$ and $\mathcal{B}=(48\pi^2)^{\frac{3\,w_\phi-1}{6\,(1+w_\phi)}}\,\lambda^{\frac{w_\phi-1}{4\,(1+\,w_\phi)}}
\mathcal{E}^{-\frac{1}{3\,(1+w_\phi)}}\,r^{\frac{1-w_\phi}{1+w_\phi}}$, with $\mathcal{E}=\frac{\epsilon\,\gamma^{-3\,w_\phi}}{2\pi\,(1+w_\phi)}$, $r=M_P/\Min$ and $\tilde{\alpha}_n=\frac{2\,(1+w_\phi)}{5-9\,w_\phi}\,\sqrt{\frac{6\,(1+w_\phi)\,(1+3\,w_\phi)}{(1-w_\phi)^2}}$.

\twocolumngrid
\bibliographystyle{apsrev}
\bibliography{Bibliography}

\end{document}